\documentclass[%
aps,
pra,
amsmath,
amssymb,
longbibliography,
reprint,
onecolumn,
superscriptaddress,
tightenlines,
12pt]{revtex4-1}
\usepackage{graphicx}
\usepackage{dcolumn}
\usepackage{bm}

\newcommand{\E}[1]{\times 10^{#1}}
\renewcommand{\vec}[1]{\bm{#1}}
\usepackage{xcolor}

\begin{document}

\author{Gaby Launay}
\email{gaby.launay@tutanota.com}
\homepage{gabylaunay.github.io}
\affiliation{
  Laboratoire de M\'ecanique des Fluides et d'Acoustique, CNRS, Universit\'e de Lyon, Ecole Centrale de Lyon, Universit\'e Lyon 1, INSA Lyon, ECL, 36 Avenue Guy de Collongue, 69134 Ecully CEDEX, France
}
\author{Tristan Cambonie}
\email{tristan.cambonie@gmail.com}
\author{Daniel Henry}
\email{daniel.henry@ec-lyon.fr}
\homepage{lmfa.ec-lyon.fr/spip.php?article368}
\affiliation{
  Laboratoire de M\'ecanique des Fluides et d'Acoustique, CNRS, Universit\'e de Lyon, Ecole Centrale de Lyon, Universit\'e Lyon 1, INSA Lyon, ECL, 36 Avenue Guy de Collongue, 69134 Ecully CEDEX, France
}
\author{Alban Poth\'erat}
\email{alban.potherat@coventry.ac.uk}
\homepage{users.complexity-coventry.org/~potherat/index.html}
\affiliation{
  Coventry University, Priory Street, Coventry, United Kingdom
}
\author{Val\'ery Botton}
\email{valery.botton@insa-lyon.fr}
\homepage{lmfa.ec-lyon.fr/spip.php?article271}
\affiliation{
  Laboratoire de M\'ecanique des Fluides et d'Acoustique, CNRS, Universit\'e de Lyon, Ecole Centrale de Lyon, Universit\'e Lyon 1, INSA Lyon, ECL, 36 Avenue Guy de Collongue, 69134 Ecully CEDEX, France
}
\affiliation{
  INSA Euro-M\'editerran\'ee, Universit\'e Euro-M\'editerran\'eenne de F\`es, Route de Mekn\`es, BP51, Fez, Morocco
}

\title{Transition to chaos in an acoustically-driven cavity flow}
\date{\today}
\begin{abstract}
  We consider the unsteady regimes of an acoustically-driven jet that forces a recirculating flow through successive reflections on the walls of a square cavity.
  The specific question being addressed is to know whether the system can sustain states of low-dimensional chaos when the acoustic intensity driving the jet is increased, and, if so, to characterise the pathway to it and the underlying physical mechanisms.
  We adopt two complementary approaches, both based on data extracted from numerical simulations:
  (i) We first characterise successive bifurcations through the analysis of leading frequencies.
  Two successive phases in the evolution of the system are singled out in this way, both leading to potentially chaotic states.
  The two phases are separated by a drastic simplification of the dynamics that immediately follows the emergence of intermittency.
  The second phase also features a second intermediate state where the dynamics is simplified due to frequency-locking.
  (ii) Nonlinear time series analysis enables us to reconstruct the attractor of the underlying dynamical system, and to calculate its correlation dimension and leading Lyapunov exponent.
  Both these quantities bring confirmation that the state preceding the dynamic simplification that initiates the second phase is chaotic.
  Poincar\'e maps further reveal that this chaotic state in fact results from a dynamic instability of the system between two non-chaotic states respectively observed at slightly lower and slightly higher acoustic forcing.
\end{abstract}

\maketitle

\section{Introduction}
\label{sec:introduction}
This paper deals with the wider issue of the pathway to chaos and turbulence in acoustically driven jet flows.
Although the possibility of driving fluid motion by means of sound waves has been known since Michael Faraday \citep{faraday_peculiar_1831}, the systematic study of these flows has only recently been tackled \citep{nyborg_acoustic_1953,lighthill_acoustic_1978}.
Whilst the basic mechanisms driving steady laminar flows are now well understood \citep{riley_steady_2001, moudjed_scaling_2014,moudjed_near-field_2015, moudjed_y-shaped_2016}, the question of their stability remains to this day a hot fundamental topic with more questions than answers \citep{dridi_influence_2008, ben_hadid_instabilities_2012, green_acoustic_2016, lyubimova_acoustic_2017, moudjed_oscillating_2014}.
Indeed, while the transition to turbulence in jets is a classical problem in fluid mechanics \citep{sato_stability_1960,landa_development_2004}, acoustically-driven jets have only recently aroused interest as a potential way of driving turbulence in fluids without the need of a direct mechanical contact.
Such technologies are crucial for the manufacture of either delicate or aggressive materials, that do not tolerate direct contact with the moving elements of a mechanical stirrer \citep{gorbunov_physical_2003}.
A typical example is the growth of crystals from a melt that is highly sensitive to impurities and more generally the stirring of a liquid during its solidification \citep{bertin_liquid-column_2010, eskin_ultrasonic_2015, oh_study_2002, dridi_influence_2008, kozhemyakin_imaging_2003, kozhemyakin_simulation_2014,chatelain_mechanical_2018}.
Though magnetic fields offer efficient solutions for contactless stirring, they demand high electric conductivities of the medium and can only act within a short distance of the walls of the fluid vessel.
By contrast, ultrasounds emitted with a suitable frequency penetrate deep into the fluid and act regardless of the electric conductivity of the fluid.
Nevertheless, the technological value of acoustic stirring would reside mostly in its ability to generate efficient mixing \citep{bulliard-sauret_heat_2017, marshall_acoustic_2015, suri_chaotic_2002}.
Since little is known of the chaotic or turbulent nature of these flows, the possibility of acoustically stirring flows in an effective way remains to be explored.

In typical configurations, an acoustic transducer is either directly inserted in the wall of a fluid vessel or placed in such a way as to generate an ultrasonic beam emitted from the wall into the fluid.
The beam propagates along a straight centerline within the vessel where fluid is accelerated, creating a jet along the beam.
Reflections may take place at the vessel walls, where two additional jets are generated: one alongside the walls and a second one driven by the reflected beam \citep{moudjed_y-shaped_2016}.
The stability properties of such a flow are to this day poorly known.
Yet, fluctuations and oscillations observed in jets forced with sufficient acoustic power suggest that a form of transition to chaos or turbulence may potentially take place \citep{lighthill_acoustic_1978, schenker_piv_2013, moudjed_oscillating_2014, cambonie_flying_2017}.
However, the nature of the transition and its underlying mechanisms remain to be found.
In particular, it is not clear whether the system sustains states of low-dimensional chaos in further stages of development of this instability, nor how much acoustic forcing is required to reach a fully turbulent state.

In an attempt to answer the first of these two questions, we consider the generic geometry of a jet driven by acoustic streaming in a square cavity.
The main jet is oriented at an angle with the walls so as to create a circulating flow pattern through successive reflections.
We numerically simulate the flow to obtain the three-dimensional, time-dependent velocity and pressure fields and tackle the question of the transition to chaos in two ways.
First, since the system sustains well-defined oscillations, it naturally lends itself to a frequency analysis, from which the emergence of chaos can be characterised by comparison to one of the classical scenarios (see \citet{mccauley_chaos_1994} for a review of these scenarios).
This method has been successful in showing that two-dimensional flows forced in a cavity, which bear resemblance with the confined flow we are considering, followed the Ruelle-Takens-Newhouse scenario \citep{molenaar_transition_2005}.
The second approach involves dynamical systems analysis based on time-series extracted from the velocity field at different locations in the flow.
This method offers a general way of characterising complex dynamical systems (see \citet{abarbanel_analysis_1993} for a review), especially low-dimensional ones.
Reconstructing the attractor of the system makes it possible to evaluate its chaotic nature.
This is done by calculating the Lyapunov exponents that characterise how quickly two initially close states of the system may diverge during their evolution through the dynamics.
The complexity of the system is further characterised by the correlation dimension of the attractor.

We first present the numerical simulations and analyse the flow regimes that are observed from a phenomenological point of view (section \ref{sec:flow-simulations}).
Section~\ref{sec:scen-trans-chaos} is dedicated to the characterisation of the flow regimes by means of frequency analysis.
Dynamical system properties are derived in section~\ref{sec:char-dynam-syst}.

\section{Flow simulations}
\label{sec:flow-simulations}
\subsection{Studied configuration}
\label{sec:stud-config}
\begin{figure}[tb]
  \centering
  \includegraphics[width=.7\columnwidth]{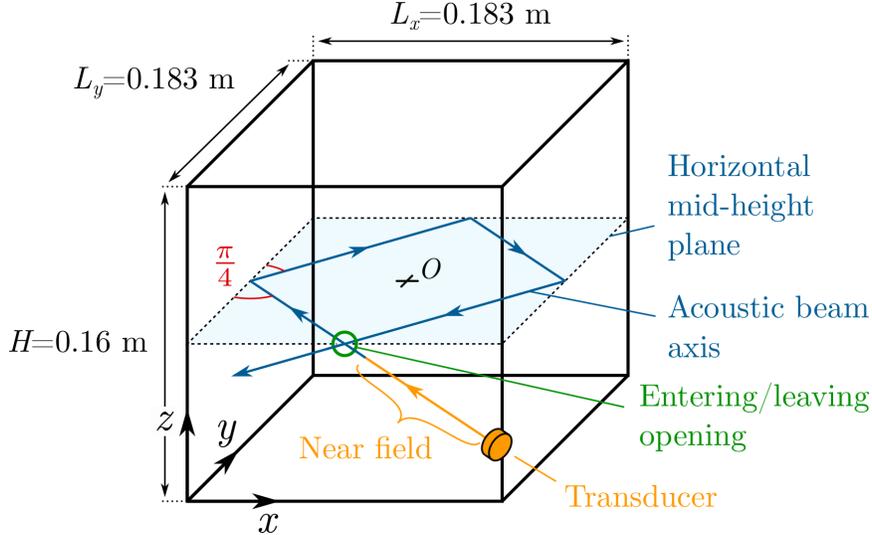}
  \caption{
    (Color online)
    Schematic representation of the cavity including the axis of the acoustic beam used to generate a circulating flow pattern.
    The acoustic beam enters the cavity by an opening in the center of one of its faces with an angle of $\pi/4$ and reflects on three of the vertical walls before leaving the cavity through the opening where it entered.
    The acoustic beam axis remains in the horizontal mid-height plane ($x, y$) and is composed of 4 rectilinear parts.
    $O$ is the origin of the reference frame used thereafter.
    This corresponds to the experimental setup used by \citet{cambonie_flying_2017}.
    The plane transducer is 28.5 mm in diameter and operates at 2 MHz.
  }
  \label{fig:schema_cavity}
\end{figure}

We study the flow created by acoustic streaming in a cavity.
The cavity is the same as in the experimental investigation by \citet{cambonie_flying_2017}:
a closed rectangular vessel of square horizontal section that is filled with water (see figure~\ref{fig:schema_cavity}).
An acoustic beam, emitted by a 2 MHz circular plane transducer of diameter 28.5 mm enters this cavity at the centre of one of its vertical sides with an angle of $\pi/4$.
As the acoustic near-field region is marbled by complex diffraction patterns and not essential to the core dynamics of the problem, it was physically separated from the investigation area to avoid unnecessary complexity.
The distance between the transducer and the cavity is then adjusted so that the acoustic beam reaches the cavity close to its far-field region \citep{moudjed_near-field_2015}.
The acoustic beam reflects successively on the three other vertical walls before leaving the cavity through the opening where it entered.
The resulting beam path is a broken line forming a square at half-height of the cavity (see figure~\ref{fig:schema_cavity}).
For the sake of simplicity, we will refer to the plane containing the acoustic beam axis as the horizontal mid-height plane.
The flow is driven by an acoustic streaming forcing in the volume inside this beam, and forms a practically square pattern inside the cavity \citep{cambonie_flying_2017}.

\subsection{Physical model}
\label{sec:phys-model}
The flow in such a configuration is governed by the Navier-Stokes equations with a force term corresponding to the acoustic forcing.
As indicated by \citet{moudjed_y-shaped_2016}, this force term can be written:
\begin{equation}
  \label{eq:1}
  \vec{F}(x, y, z) = \frac{2 \alpha I_{ac}(x, y, z)}{c} \, \vec{e}_p,
\end{equation}

where $\vec{F}$ is the force per unit of volume (kg.m$^{-2}$.s$^{-2}$), $\alpha$ the acoustic attenuation (m$^{-1}$), $I_{ac}$ the acoustic intensity (W.m$^{-2}$), $c$ the sound velocity (m.s$^{-1}$), and $\vec{e}_p$ the acoustic beam direction of propagation.
$I_{ac}$ can be further expressed as $I_{ac}=I_0 \, I(x, y, z)$, with $I_0$ the acoustic intensity amplitude (W.m$^{-2}$) and $I(x, y, z)$ the normalised acoustic intensity distribution.
According to \citet{blackstock_fundamentals_2000}, the acoustic intensity amplitude $I_0$ is directly related to the transducer acoustic power $P_{ac}$ (W) by $I_0=P_{ac}/(\pi R^2_t/4)$, where $R_t$ is the transducer radius (m).
The acoustic force can then be written as:
\begin{equation}
  \label{eq:2}
  \vec{F}(x, y, z) = \left[ \frac{2 \alpha P_{ac}}{c \pi R_t^2} \right] 4 I(x, y, z) \, \vec{e}_p.
\end{equation}

The problem is made dimensionless using the cavity height $H$ as length scale, $\nu/H$ as velocity scale, $H^2/\nu$ as time scale, and $\rho\nu^2/H^2$ as pressure scale, with $\rho$ the fluid density and $\nu$ the kinematic viscosity.
The dimensionless equations to solve in the cavity are then

\begin{equation}
  \nabla \cdot \vec{u} = 0,
  \label{eq:3}
\end{equation}
\begin{equation}
  \frac{\partial \vec{u}}{\partial t} + \left(\vec{u}\cdot\nabla \right) \vec{u} =  -\nabla p + \nabla^2 \vec{u}
  + A^* 4 \, I(x, y, z) \, \vec{e}_p,
  \label{eq:4}
\end{equation}
with $A^*$, the dimensionless magnitude of the acoustic forcing, given by
\begin{equation}
  \label{eq:5}
  A^* = \frac{2 \alpha P_{ac} H^3}{c \pi R_t^2 \rho \nu^2} \, .
\end{equation}
Instead of $A^*$, we shall use the quantity $A=10^{-6}A^*$ as the sole control parameter. With this choice, $A=1$, corresponds to a dimensional power $P_{ac}$ of $1$ W, for the parameters of \citep{cambonie_flying_2017}'s experiment (\textit{i.e.} $\alpha=0.1$ m$^{-1}$ for a 2 MHz transducer, $\rho=10^3$~Kg.m$^{-3}$, $c=1480$~m.s$^{-1}$, $\nu=9.3 \;10^{-7}$~m$^2$/s, $R_t=0.01425$~m and $H=0.16$~m)

\begin{figure}[tb]
  \centering
  \includegraphics[width=.65\columnwidth]{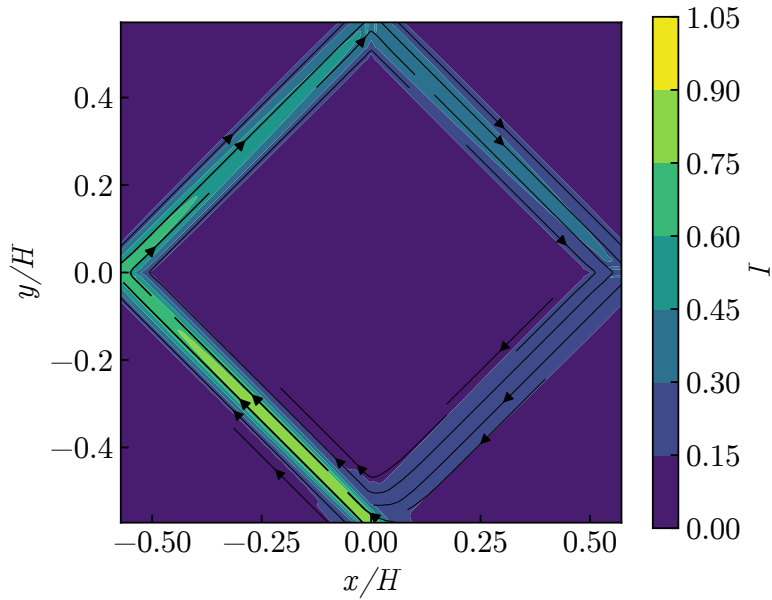}
  \caption{
    (Color online)
    Normalized acoustic forcing $I$ generated by the acoustic beam in the horizontal mid-height plane ($z/H=0$), computed using the Rayleigh integral.
    The acoustic beam enters at ($x/H=0$, $y/H=-0.571875$), the center of a vertical wall, with a $\pi/4$ angle and is reflected on the other vertical walls while slowly diverging and dissipating (which explains the progressive loss of intensity), before leaving the cavity at its inlet location.
  }
  \label{fig:Forcing}
\end{figure}

The normalised acoustic intensity distribution $I(x, y, z)$ in the cavity is computed for the above parameters by means of the Rayleigh integral for each beam \citep{blackstock_fundamentals_2000,moudjed_y-shaped_2016}.
These beams come from virtual sources obtained as images of the acoustic source through successive symmetries with respect to the different walls where the reflections occur.
The progressive attenuation of the acoustic beam along its propagation path is taken into account.
More details on diffraction, attenuation and reflection at the walls can be found in \citet{moudjed_y-shaped_2016}.
In particular, these authors have shown that the treatment of the zones where incident and reflected beams interfere is not crucial to reliably derive the flow pattern.
For each reflection zone, we have then chosen to use the incident field up to the vertical normal plane and the reflected beam beyond this plane.
These results are presented in figure~\ref{fig:Forcing}.
Note that, due to the divergence and dissipation of the acoustic beam, the four branches of the forcing are not identical.
Consequently, the only symmetry of the forcing is about the $z=0$ plane.

\subsection{Numerical simulations}
\label{sec:numer-simul}
The simulations of the flow driven by this steady acoustic forcing are run using a spectral finite element method \citep{hadid_numerical_1997}, with a grid comprising two elements in the vertical direction (along $z$, with 21 points per element) and two elements in each of the two horizontal directions (with 31 points per element).
In each element, the spatial discretization is obtained through Gauss-Lobatto-Legendre points distributions which are naturally tightened along the element boundaries.
The time discretization is carried out using a semi-implicit splitting scheme, as proposed by \citet{karniadakis_1991}: the nonlinear terms are first integrated explicitly, the pressure is then solved through a pressure equation enforcing the incompressibility constraint (with a consistent pressure boundary condition derived from the equations of motion), and the linear terms are finally integrated implicitly.
This time integration scheme is used throughout our numerical simulations with the third-order accurate formulation described in \citet{karniadakis_1991}.
A no-slip boundary condition is applied at all the cavity walls.

In our configuration, the acoustic forcing principally occurs in the neighbourhood of the horizontal mid-plane.
In order to accurately take into account this forcing, which is imposed at the grid points, we use a multi-element spectral approach allowing us to choose two elements in the vertical direction.
The mesh is then refined at mid-height with 13 discretization points on the width of the acoustic intensity peak and jet velocity peak.
Tests with different meshes have ensured that this choice provides a good discretization of the imposed acoustic forcing and a good precision for the calculation of the main flow which also occurs in this region at mid-height of the cavity.%
Figure \ref{fig:topo_evolution} presents a comparison of the experimental and numerical time-averaged and RMS velocity fields in the horizontal mid-height plane.
In the experimental case, the relation between the acoustic power ($P$) and the acoustic forcing ($A$) is not known, making \emph{a priori} comparison difficult.
Instead, cases that, as best as possible with the available data, exhibit the same typical values of velocity \emph{a posteriori} are presented.
The numerical simulations reproduce well the time-averaged topology of the experimental flow (figure~\ref{fig:topo_evolution}a-d), featuring the 4 successive jets and their associated wall-jets.
Regarding the RMS of the velocity (figure~\ref{fig:topo_evolution}g-j), the flow behaviours are qualitatively similar albeit with more localised peaks of intensity in the zones where the beam reflects on the wall in the experiments.
Nevertheless, in the absence of sufficient control of the acoustic force in the experiment, it is difficult to push the comparison further than the qualitative level.

\subsection{Flow overview}
\label{sec:flow-overview}
To investigate this flow, we first present its global evolution with the acoustic forcing $A$ (mean flow and fluctuation distribution).
We then deepen our analysis by considering the velocity time-series at specific locations and for different values of $A$.

\begin{figure}[bt]
  \centering
  \includegraphics[width=.9\columnwidth]{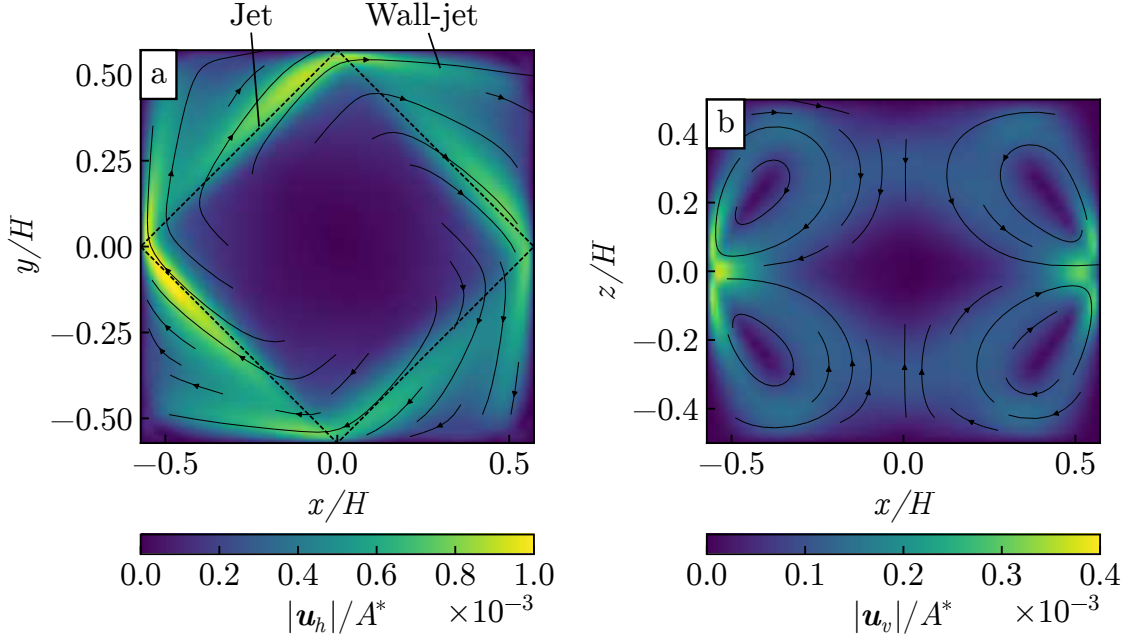}
  \caption{
    (Color online)
    Normalized velocity fields for a steady configuration ($A=1$).
    (a)~inplane velocity magnitude $|\vec{u}_h|/A^*$ and segments of 2D streamlines in the horizontal mid-height plane ($z=0$).
    Dashed lines represent the acoustic beam axis.
    (b)~inplane velocity magnitude $|\vec{u}_v|/A^*$ and segments of 2D streamlines in the vertical plane ($y=0$).
  }
  \label{fig:velocity_fields_A1_A15}
\end{figure}

For low values of the acoustic forcing ($A<1.5$), the system is steady.
An example of steady state flow is presented in figure~\ref{fig:velocity_fields_A1_A15} for $A=1$ through the values of the velocity norm at mid-height and in the $y=0$ vertical plane.
For moderately low acoustic forcing ($A>0.5$, see \citet{cambonie_flying_2017}), a jet reaching a wall generates
(i) a wall jet by inertial effect, which remains in the vicinity of the wall (see figure~\ref{fig:velocity_fields_A1_A15}a), and
(ii) vertical velocities, which create recirculation zones above and below the horizontal mid-height plane (see figure~\ref{fig:velocity_fields_A1_A15}b).
This flow features two quasi-symmetries, a 4-quadrant quasi-symmetry around the $z$ axis (visible in figure~\ref{fig:velocity_fields_A1_A15}a), and a symmetry with respect to the horizontal mid-height plane (up-down symmetry visible in figure~\ref{fig:velocity_fields_A1_A15}b).

\begin{figure}[tb]
  \centering
  \includegraphics[width=\columnwidth]{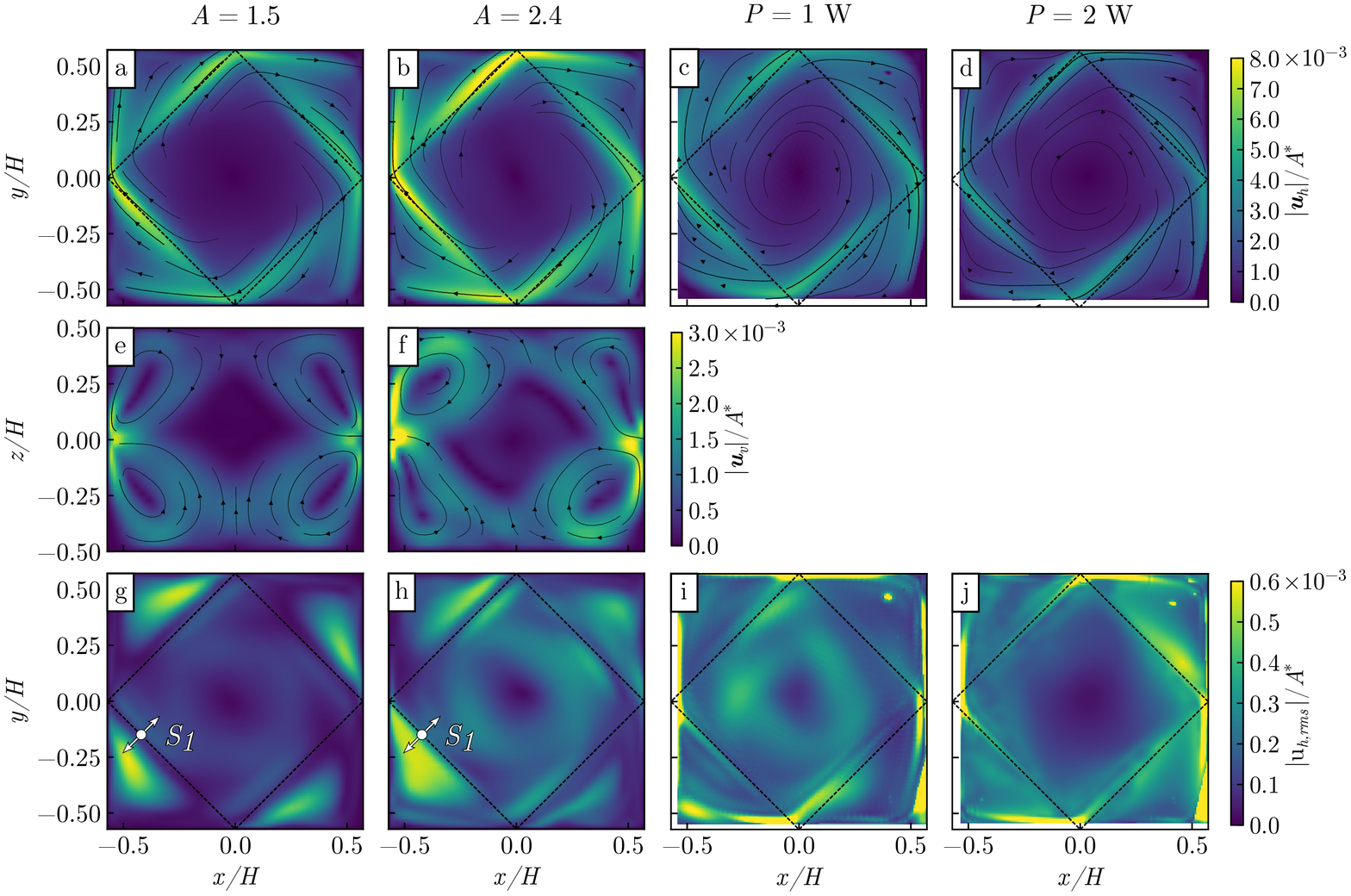}
  \caption{
      (Color online)
      Normalized time-averaged velocity fields and RMS fluctuations for various acoustic forcing.
      (a-d):~inplane velocity magnitude $|\overline{\vec{u}_h}|/A^*$ and segments of 2D streamlines in the horizontal mid-height plane ($z/H=0$).
      (a) and (b):~Numerical results for $A=1.5$ and $A=2.4$.
      Dashed lines represent the acoustic beam axis.
      (c) and (d):~Experimental PIV measurements for $P=1$~W and $P=2$~W.
      The velocity fields are normalized using respectively $A^*=1.5\E{6}$ and $A^*=2.4\E{6}$ for the sake of the comparison with the numerical results.
      Dashed lines represent the expected position of the acoustic beam.
      (e) and (f):~inplane velocity magnitude $|\overline{\vec{u}_v}|/A^*$ and segments of 2D streamlines in the vertical plane ($y=0$) for $A=1.5$ and $A=2.4$.
      (g) and (h):~normalized RMS fluctuations $|\vec{u}_{h, rms}|/A^*$ in the horizontal mid-height plane for $A=1.5$ and $A=2.4$.
      $S_1$ is the point where velocity time-series are recorded for subsequent spectral and non-linear dynamics analyses.
      Arrows on this point indicate the transverse direction.
      (i) and (j):~normalized experimental RMS fluctuations $|\vec{u}_{h, rms}|/A^*$ in the horizontal mid-height plane for $P=1$~W and $P=2$~W.
    }
  \label{fig:topo_evolution}
\end{figure}

For $A=1.5$, the system becomes periodic, with velocity fluctuations concentrated between the impinging jets and the walls (see figure~\ref{fig:topo_evolution}g).
The dimensionless period associated to this oscillation $T=7.03\times10^{-3}$ is of the same order as the period measured by \citet{cambonie_flying_2017} ($T=6.40\times10^{-3}$ for $A \approx 1$).
Oscillations featured in this configuration are strongly 3D and complex in structure.
This makes it challenging to identify the instability generating them.
However, the topology of the time-averaged velocity field (in both horizontal and vertical planes, respectively in figures \ref{fig:topo_evolution}a and e) remains very similar to that for $A=1$, and the distribution of the RMS velocity fluctuations shares the 4-quadrant quasi-symmetry of the time-averaged flow (see figure~\ref{fig:topo_evolution}g).

\begin{figure}[tb]
  \centering
  \includegraphics[width=.7\columnwidth]{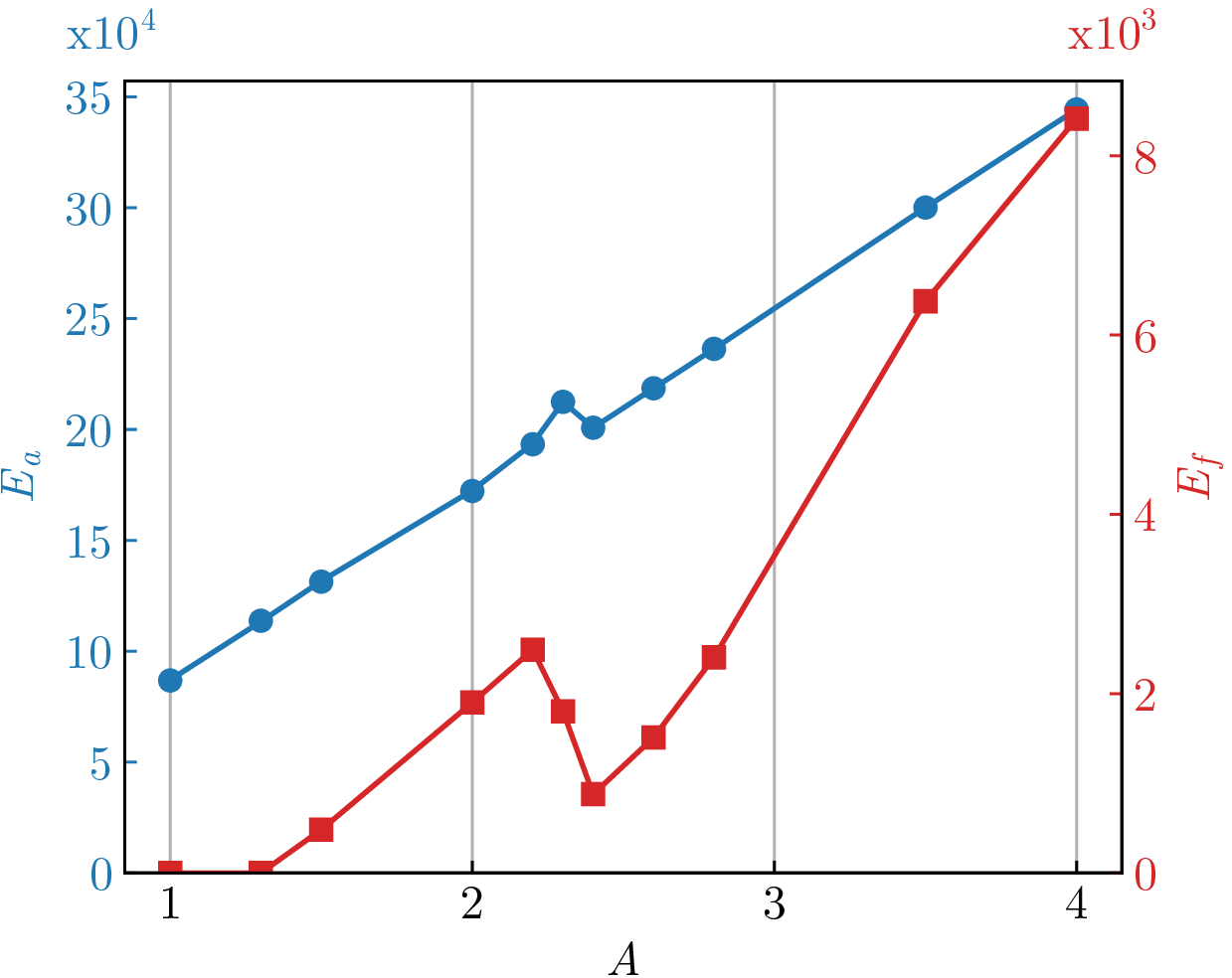}
  \caption{
    (Color online)
    Evolution of the cavity flow energy with increasing acoustic forcing $A$.
    Blue circles: time-averaged energy $E_a$ in the cavity, from equation (\ref{eq:6}).
    Red squares: fluctuation energy $E_f$ in the cavity, from equation (\ref{eq:7}).
  }
  \label{fig:NRJ_evolution}
\end{figure}

For acoustic forcing between $A=1.5$ and $A=2.2$, the time-averaged flow topology remains very similar (not shown here).
In particular, it conserves the up-down and 4-quadrant quasi-symmetries
However, the energy of both the time-averaged flow and the fluctuations increases, as shown in figure~\ref{fig:NRJ_evolution}, as a result of the higher forcing.
The energy values presented in this figure are obtained by computing the integral of the dimensionless kinetic energy over the cavity volume $V$:
\begin{equation}
  E_{a} = \frac{1}{2} \int_V |\overline{\vec{u}}|^2 \, \mathrm{d}V,
  \label{eq:6}
\end{equation}
\begin{equation}
  E_{f} = \frac{1}{2} \int_V u_{rms}^2\,\mathrm{d}V.
  \label{eq:7}
\end{equation}
The time-averaged and RMS velocity fields are defined as:
\begin{equation}
\overline{\vec{u}} = \frac{1}{t_f} \int_{t=0}^{t=t_f} \vec{u}\,\mathrm{d}t,
  \label{eq:6b}
\end{equation}
\begin{equation}
  u_{rms} = \left[ \frac{1}{t_f} \int_{t=0}^{t=t_f} | \vec{u} - \overline{\vec{u}} |^2\,  \mathrm{d}t \right]^{\frac{1}{2}},
  \label{eq:7b}
\end{equation}
with $t_f$ the duration of the simulation.

For $A=2.4$, both vertical and horizontal symmetries are broken:
while the time-averaged flow in the horizontal mid-height plane (figure~\ref{fig:topo_evolution}b) remains similar to that for $A=1.5$, the time-averaged flow in the vertical plane (figure~\ref{fig:topo_evolution}f) and the fluctuations in the horizontal mid-height plane (figure~\ref{fig:topo_evolution}h) lose symmetry.
This reflects a major change in the flow dynamics, that is confirmed by the sudden decrease in the fluctuation energy $E_f$ for $A=2.4$ (figure~\ref{fig:NRJ_evolution}).

For an acoustic forcing in the range $A=2.4$ to $A=4$, the time-averaged and fluctuation energies keep increasing (see figure~\ref{fig:NRJ_evolution}), but neither the topology of the average flow, nor that of the veclocity fluctuations evolve significantly.
It is noteworthy that \citet{cambonie_flying_2017} observed more significant alterations of the time-averaged flow topology in the mid-height horizontal plane for a wider range of acoustic forcing ($A \in [0.5, 8]$).

\begin{figure}[p]
  \centering
  \includegraphics[width=.9\columnwidth]{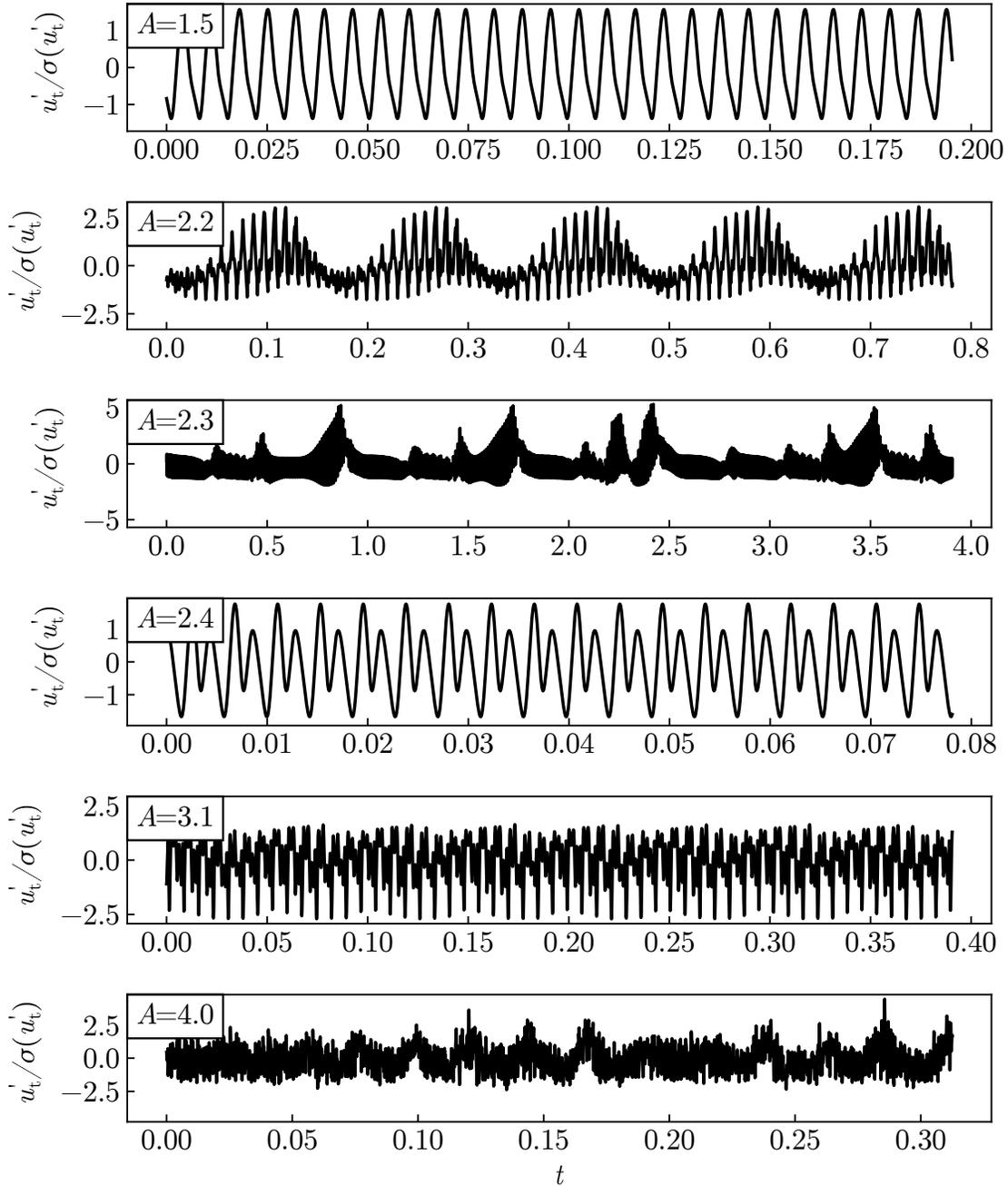}
  \caption{
    Temporal evolution of the jet transverse velocity fluctuation ($u_t'$) for increasing acoustic forcing $A$.
    The transverse velocity fluctuation $u_t'$ is the fluctuating part of the velocity component perpendicular to the acoustic axis, in the horizontal mid-height plane at point $S_1$ (see point $S_1$ and associated arrows in figure~\ref{fig:topo_evolution}g and h).
    The velocity is normalized by the standard deviation of the velocity fluctuations $\sigma(u'_t)$.
    Note that the durations observed on these plots correspond to several hours in physical time ($5 \times 10^3$ s to $10^5$ s).
    It is worth stressing that such long time-series are indispensible to adequately represent the system's dynamics.
    However, experimentally obtaining data over such long periods of time would be extremely difficult.
  }
  \label{fig:time-series-evolution}
\end{figure}

The configuration for $A=2.3$ appears to be transitional in terms of energy evolution (see figure~\ref{fig:NRJ_evolution}), but still exhibits the vertical and horizontal symmetries seen at lower acoustic forcing.
In contrast to the relatively gentle changes in the mean and fluctuation fields within the ranges $A=1.5$ to $A=2.3$ and $A=2.4$ to $A=4$, the time-series of the jet velocity is significantly altered within these same ranges (see figure~\ref{fig:time-series-evolution}).
More specifically,
(i) The fluctuations change from almost sinusoidal to quasi-periodic between $A=1.5$ and $A=2.2$.
(ii) For $A=2.3$, the velocity fluctuations are rather complex, with no apparent periodicity.
(iii) At $A=2.4$, the time-series again adopts a periodic behavior.
(iv) Between $A=2.4$ and $A=4.0$, the series progressively loses any apparent periodicity or structure.

The non-trivial succession of periodic, quasi-periodic, periodic again and eventually seemingly random regimes is reminiscent of classical transition scenarios to chaos.
It occurs over a range of parameters that is consistent with the experimental observations of \citet{cambonie_flying_2017}.

\section{Scenario for the transition to chaos}
\label{sec:scen-trans-chaos}
The sequence of flow regimes observed in the fluctuations (figure~\ref{fig:topo_evolution}) and time-series (figure~\ref{fig:time-series-evolution}) shows that overall, the flow transits to more chaotic states as the forcing parameter $A$ is increased.

\begin{figure}[tb]
  \centering
  \includegraphics[width=.9\columnwidth]{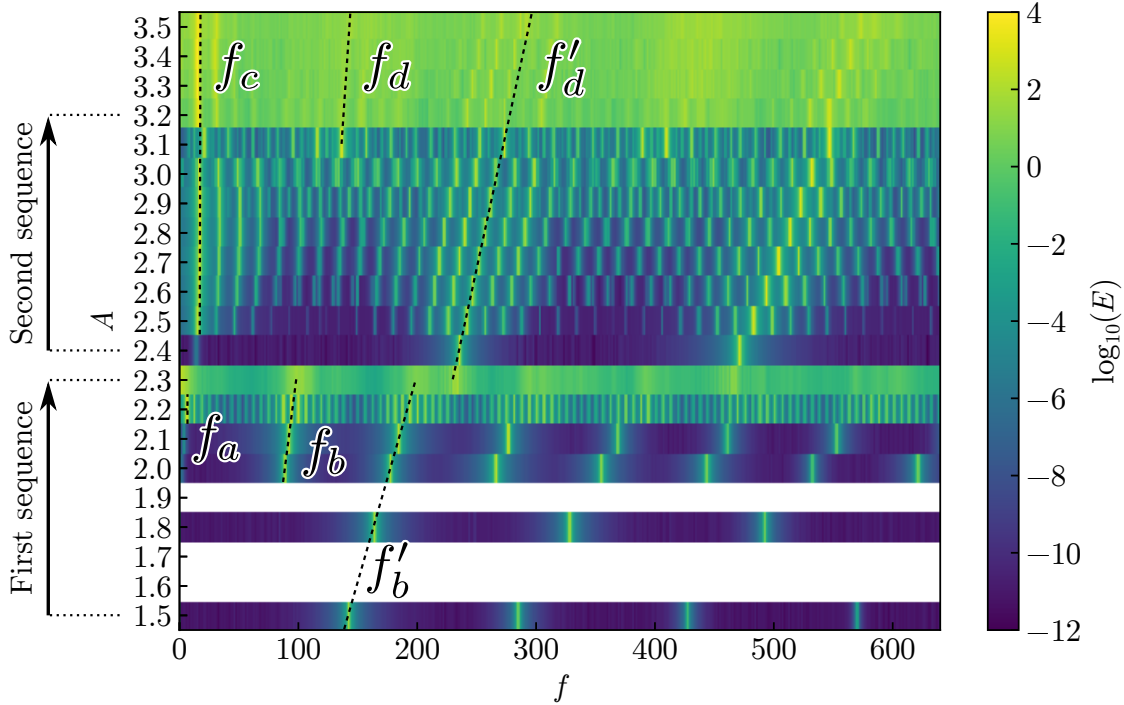}
  \caption{
    (Color online)
    Power spectral density $E$ as a function of the acoustic forcing $A$ and the frequency $f$.
    $E$ is computed using fast Fourier transform of the time-series of $u_t'$ following Welch's method \citep{welch_use_1967}.
    This picture singles out the appearence of new frequencies associated with bifurcations (Hopf and period-doubling) and highlights potentially chaotic regimes (at $A=2.3$ and $A>3.2$).
    Dashed lines identify particular frequencies evolving with $A$.
    Note that the observed range of frequency values is consistent with the very low frequencies experimentally observed by \citet{cambonie_flying_2017} (of the order of $5\E{-3}$Hz, or $f=128$).
  }
  \label{fig:spectrum_evolution}
\end{figure}

This evolution, however, presents unusual features, such as the brutal change in flow topology and fluctuation energy that occurs between $A=2.2$ and $A=2.4$ (see figure~\ref{fig:NRJ_evolution}).
Classically, the route to chaos follows one of the three canonical scenarios:
(i) Successive appearance of low frequencies (Hopf bifurcations) are observed in the Ruelle-Takens-Newhouse scenario \citep{newhouse_occurrence_1978}.
In this scenario, the chaotic behavior typically appears after three bifurcations \citep{eckmann_roads_1981}.
(ii) Succession of period-doubling bifurcations are part of the Feigenbaum scenario \citep{feigenbaum_quantitative_1978, feigenbaum_onset_1979, feigenbaum_transition_1980}.
In this scenario, chaos can appear after a very large number of bifurcations of this type.
(iii) Appearance of intermittent chaotic events that occupy an increasingly large fraction of the timeline (intermittency phenomenon) are observed in the Pomeau-Manneville scenario \citep{pomeau_intermittent_1980}.
The characteristic features of each of these scenarios can be tracked in the evolution of the dominating frequencies of the system, and through the possible occurrence of intermittent behaviour (see for example \citep{mccauley_chaos_1994}).
Following this idea, we shall now analyze the main frequencies of velocity fluctuations and seek similarities with the classical scenarios for the transition to chaos.
For this we will focus on $u'_t$, the fluctuation of the horizontal velocity component transverse to the first beam axis at point $S_1$ (see figure~\ref{fig:topo_evolution}g and h).
Figure~\ref{fig:spectrum_evolution} shows the power spectral density $E$ for different values of the acoustic forcing $A$ extracted from the time series of $u'_t$.
It has been verified that the peak frequencies appearing in the power spectral densities are independent of the position in space within the cavity, as long as they are extracted from an area with significant fluctuation intensity.

\subsection{Spectral signature of the different regimes}
\label{sec:spectr-sign-diff}
For the two lowest acoustic forcings ($A=1.5$ and $A=1.8$), the flow is periodic and the power spectral densities show a peak frequency (denoted $f_b'$ in figure~\ref{fig:spectrum_evolution}, at $f=143$ and $f=164$, respectively for $A=1.5$ and $A=1.8$) along with harmonics at higher frequencies.
These very low frequencies (on the order of $5\E{-3}$Hz) are consistent with the experimental observations by \citet{cambonie_flying_2017}.
The peak frequency increases linearly with $A$ and can be traced up to $A=2.3$.
From $A=2.0$, a peak at half frequency $f_b=0.5f_b'$ appears, indicating a period-doubling.
$f_b$ also increases with $A$ and exists up to $A=2.3$.
Successive period-doublings is one of the three identified scenarios of transition to chaos \citep{eckmann_roads_1981}.
In the present case however, the next peak, which appears for an acoustic forcing $A=2.2$, is at a far lower frequency $f_a=7$.
$f_a$ is not commensurate with $f_b$ and results from a Hopf bifurcation (as defined by \citet{eckmann_roads_1981}).
The harmonics of this new fundamental frequency are visible across the entire resolved spectrum.
In particular, they interfere with the peak frequencies associated with $f_b$ and $f_b'$.

\begin{figure}[p]
  \centering
  \includegraphics[width=.9\columnwidth]{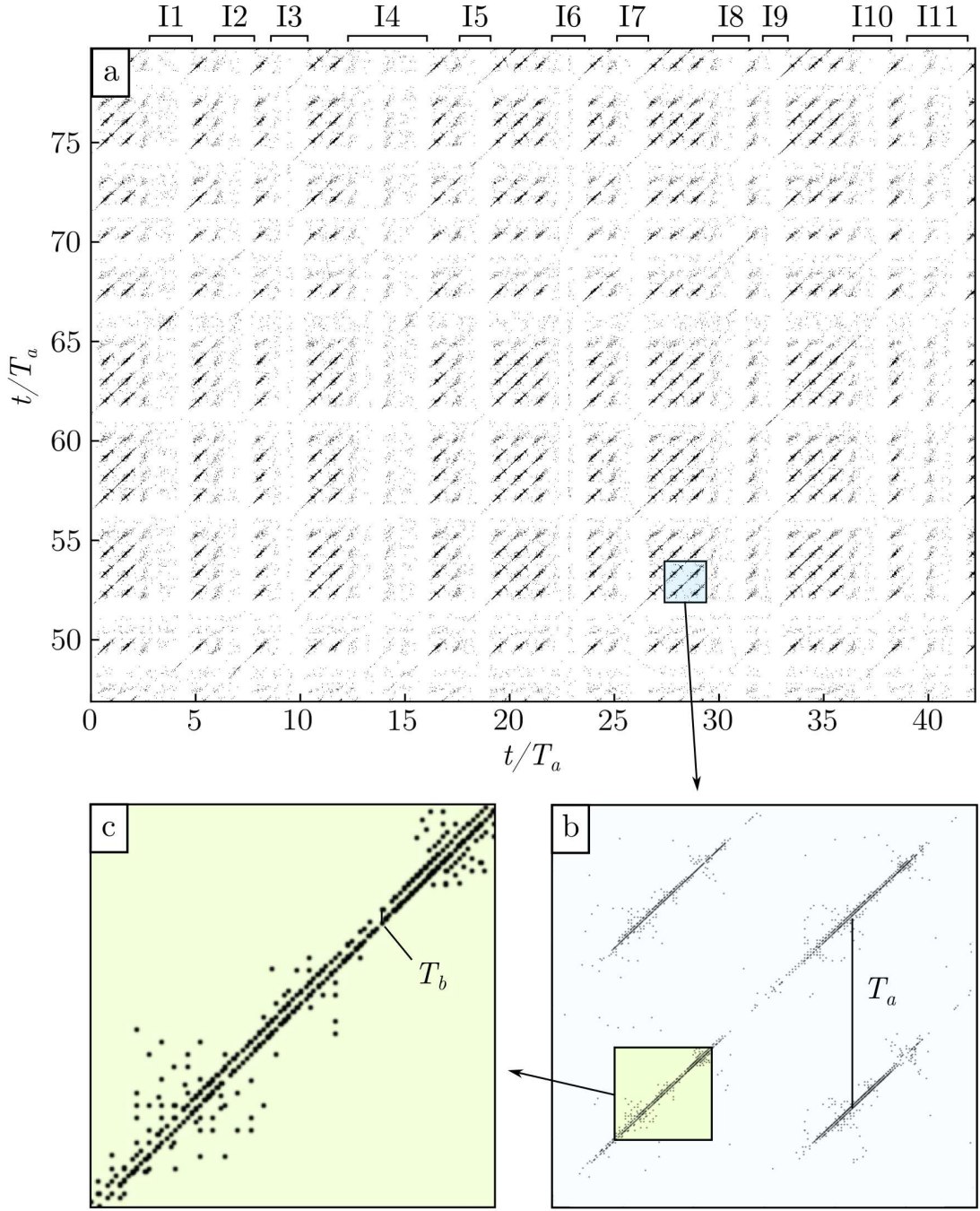}
  \caption{
    (a) Recurrence map of $u_t'$ for acoustic forcing $A=2.3$.
    The regions with diagonal patterns correspond to time intervals for which the system is quasi-periodic.
    White horizontal and vertical bands correspond to intermittencies, for which the dynamical system's evolution looses quasi-periodicity.
    For the sake of simplicity, only a portion of the signal is represented here, where only 11 of the 25 intervals of intermittent behaviour ($I1$ to $I11$) from the full measured signal are visible.
    (b) and (c): enlargements of the recurrence map.
    $T_a$ and $T_b$ are the periods associated with the characteristic frequencies $f_a$ and $f_b$ on figure~\ref{fig:spectrum_evolution}.
  }
  \label{fig:recurr_map_2_3}
\end{figure}

\begin{figure}[tb]
  \centering
  \includegraphics[width=.7\columnwidth]{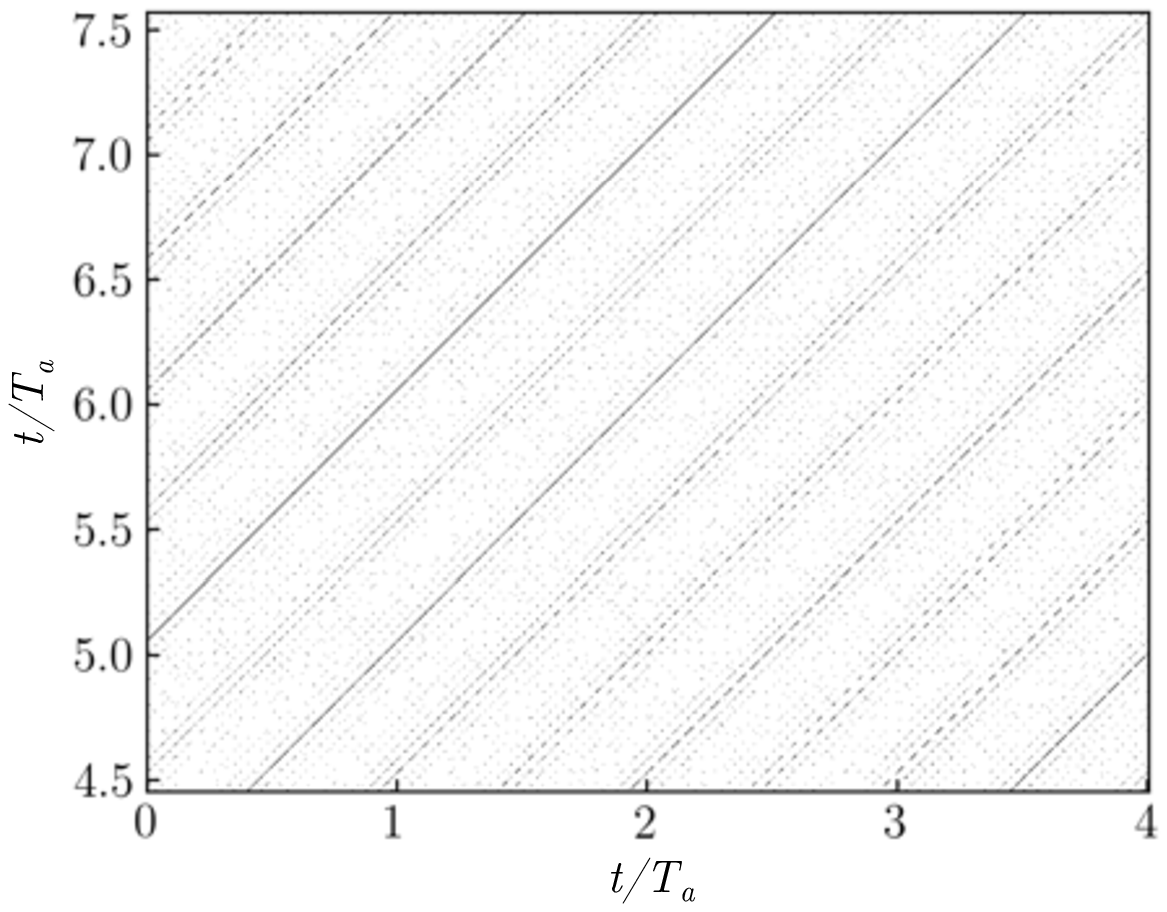}
  \caption{
    Recurrence map of $u_t'$ for an acoustic forcing $A=2.2$.
    The diagonal patterns indicating a quasi-periodic behavior are clearly visible on this figure.
  }
  \label{fig:recurr_map_2_2}
\end{figure}

The next dynamical change occurs at $A=2.3$, where the energy suddenly spreads across the spectrum.
This can be seen as the floor level of the spectrum raises from noise level (about $10^{-10}$) at $A=2.2$ to $10^{-3}$ at $A=2.3$.
This opens the door to a possible intermittent behaviour, which is indeed detected by means of the recurrence map shown in figure~\ref{fig:recurr_map_2_3}.
On this map, each black dot on a recurrence represents two moments in time (respectively on abscissa and ordinate) for which the system states are close in the phase space (here, two moments where $u_t'$ values differ by less than $0.1\%$ of the total velocity fluctuation amplitude).
Consequently, diagonal lines formed by succession of dots represent time intervals for which the signal is correlated.
Hence, periodic signals show up as diagonal lines distant by the signal period.
For $A=2.3$, a portion of the recurrence map computed in a phase space of dimension $3$ is presented in figure~\ref{fig:recurr_map_2_3}a.
This map exhibits regions of quasi-periodicity (diagonal patterns), separated by horizontal and vertical white bands (marked as $I1$ to $I11$), for which the dynamical system exhibits a behavior where quasi-periodicity is lost. This is the signature of intermittencies.
For comparison, a recurrence map that features no intermittencies ($A=2.2$) is shown in figure~\ref{fig:recurr_map_2_2}.
Zooming on a region with diagonal patterns (figures \ref{fig:recurr_map_2_3}b-c) reveals the frequencies of the quasi-periodic signal: $f_a=1/T_a=6.92$ and $f_b=1/T_b=113$, in agreement with the frequencies identified on the power spectral density (figure~\ref{fig:spectrum_evolution}).


For $A=2.4$, the power spectral densities reflect a drastic simplification of the dynamics, which reverts to being periodic.
This simplification coincides with the drop of fluctuation energy observed in figure~\ref{fig:NRJ_evolution}.
The new fundamental frequency at $A=2.4$ reads $f_d'=236$, increases linearly with $A$ and exists up to $A=3.5$.
At $A=2.5$, a Hopf bifurcation gives birth to a low frequency $f_c=15.9$, similarly to what occurs at $A=2.2$.
At $A=3.1$, a period-doubling bifurcation produces a new frequency $f_d=f_d'/2=137$, as well as a substantial shift of the frequency $f_c$ towards a higher value.
Finally, for $A\geq 3.2$ (including $A=4$, not shown here), the energy spreads across the spectrum, as expected for chaotic systems.
Nevertheless, the peak frequencies exhibited at $A=3.1$ can still be traced in the continuous spectra.

\subsection{Evolution of the frequencies with the forcing parameter A and frequency locking}
\label{sec:char-freq}
\begin{figure}[tb]
  \centering
  \includegraphics[width=.7\columnwidth]{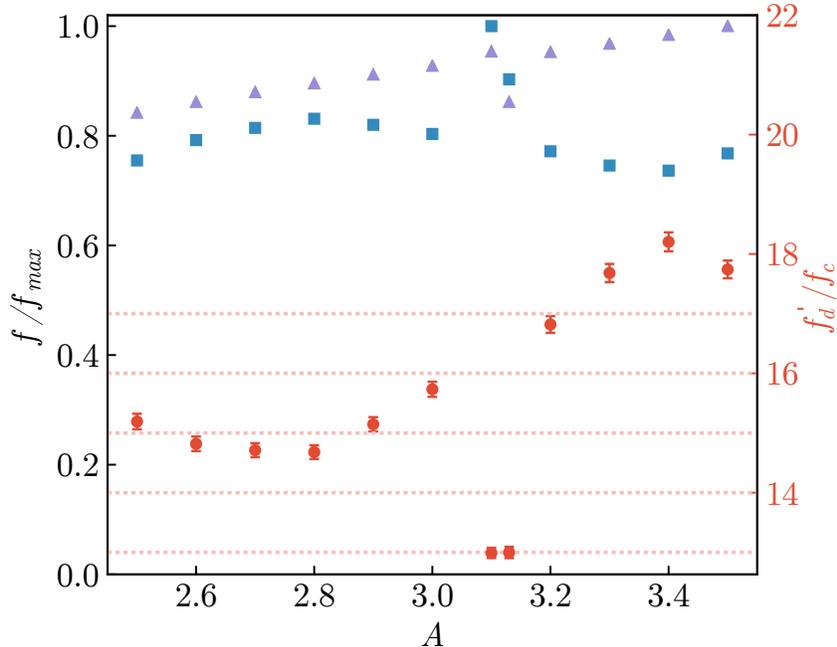}
  \caption{
    (Color online)
    Evolution of characteristic frequencies with the acoustic forcing $A$.
    Blue squares: normalized frequency $f_c$.
    Purple triangles: normalized frequency $f'_d$.
    Red circles: ratio $f'_d/f_c$ (right axis).
    The peak frequencies are obtained from local maxima in the power spectral densities, and hence bear an uncertainty $\Delta f/f_{max} = 1/(t_f f_{max}) < 4.5\E{-4}$ linked to the times-series duration $t_f$.
    The frequency $f_c$ appears to evolve smoothly, with the exception of the cases between $A=3.1$ and $A=3.13$, where it is abnormally high.
    The fact that the ratio $f_d'/f_c$ is an integer for these cases suggests that this abnormal behaviour is due to frequency-locking.
  }
  \label{fig:low_frequency_evolution}
\end{figure}

The evolution of the frequencies identified above gives a first indication on the underlying flow dynamics.
We have noted that the high frequencies (namely $f_b$, $f_d$ and their harmonics) evolve linearly with the acoustic forcing $A$ (see figure~\ref{fig:spectrum_evolution}).
The low frequency $f_c$, in contrast, does not evolve monotonically (figure~\ref{fig:low_frequency_evolution}).
All of these variations however remain smooth, with the notable exception of configurations with acoustic forcing between $A=3.1$ and $A=3.13$, where $f_c$ suddenly shifts to a higher frequency and $f_d$ also undergoes little discontinuities.

These first discontinuities may be understood by recalling that when two oscillating phenomena coexist in a single dynamical system, the spectra exhibit frequency peaks at their linear combinations (explaining the numerous peaks observed at $A=2.2$ and in the range $A \in [2.5, 3.1]$ in figure \ref{fig:spectrum_evolution}).
However, if the ratio between these two frequencies happens to be a rational number, the dynamics can be drastically simplified.
This effect is called \emph{frequency-locking} \citep{mccauley_chaos_1994}.
For these specific values, one can expect much simpler dynamics, compared to neighbouring values of $A$.
This is for example the case for configurations with acoustic forcing between $A=3.1$ and $A=3.13$, where $f_d/f_c=T_c/T_d=13/2$, meaning that the oscillation periods associated to $f_c$ and $f_d$ are in a ratio of $13$ to $2$.
This simplification is clearly visible on the evolution of the dimension of the system, which will be presented in section~\ref{sec:embedd-corr-dimens}.
The existence of frequency-locking on an interval of the governing parameter is known in non-linear dynamical system theory as ``Arnold's tongue'' \citep{mccauley_chaos_1994}.

The evolution of the main frequencies displays a second set of discontinuities when the dynamics suddenly simplify at $A=2.4$: although $f_d'$, $f_d$ and $f_c$ are respectively close to $f_b'$, $f_b$ and $f_a$, their evolution exhibits a discontinuity in the range $A=2.3$ to $A=2.4$ (see figure~\ref{fig:spectrum_evolution}).
We shall analyze this phenomenon more in detail in section~\ref{sec:pointc-sect-symm}.

\subsection{Scenario for the transition to chaos}
\label{sec:trans-chaos-scen-1}
The evolution of our system exhibits elements of all three canonical scenarios of transition to chaos (Ruelle-Takens-Newhouse, Feigenbaum and Pomeau-Manneville) but complies fully with none of them.
A potentially chaotic regime first appears for $A=2.3$, where intermittency is observed.
It follows a first Hopf bifurcation to the periodic basic state and then a period-doubling bifurcation and a Hopf bifurcation.
After the dynamics has simplified back to a periodic state, another two bifurcations are needed for chaos to appear again.
This time, the period-doubling and the Hopf bifurcations appear in reverse order.
The most striking difference with the three established scenarios of transition to chaos is the sudden simplification of the dynamics that occurs at $A=2.4$, which implies that two distinct ranges of the forcing parameter can potentially lead to chaos.
These two distinct ranges will be referred to as the first and second sequences in the sequel.
This peculiar feature raises two questions:
first, do both ranges of regimes where a continuous frequency spectrum is observed support actual chaos ?
And second, which mechanisms lead to the simplification observed at $A$=2.4 ?
We shall now attempt to answer these questions.

\section{Characterisation of the dynamical system}
\label{sec:char-dynam-syst}
Chaotic dynamical systems cannot be fully characterized by means of frequency spectra.
In particular, spectra do not offer a way to distinguish stochastic systems from chaotic but still deterministic ones, as both exhibit continuous spectra.
To ascertain the possible chaotic nature of the system, we shall now seek to characterize the underlying dynamical system by means of non-linear time-series analyses.
More specifically, we shall seek the conditions in which the dimension of the dynamical system becomes fractal, and when sensitivity to initial conditions, as measured by Lyapunov exponents betrays a chaotic behaviour.

\begin{figure}[tb]
  \centering
  \includegraphics[width=.7\columnwidth]{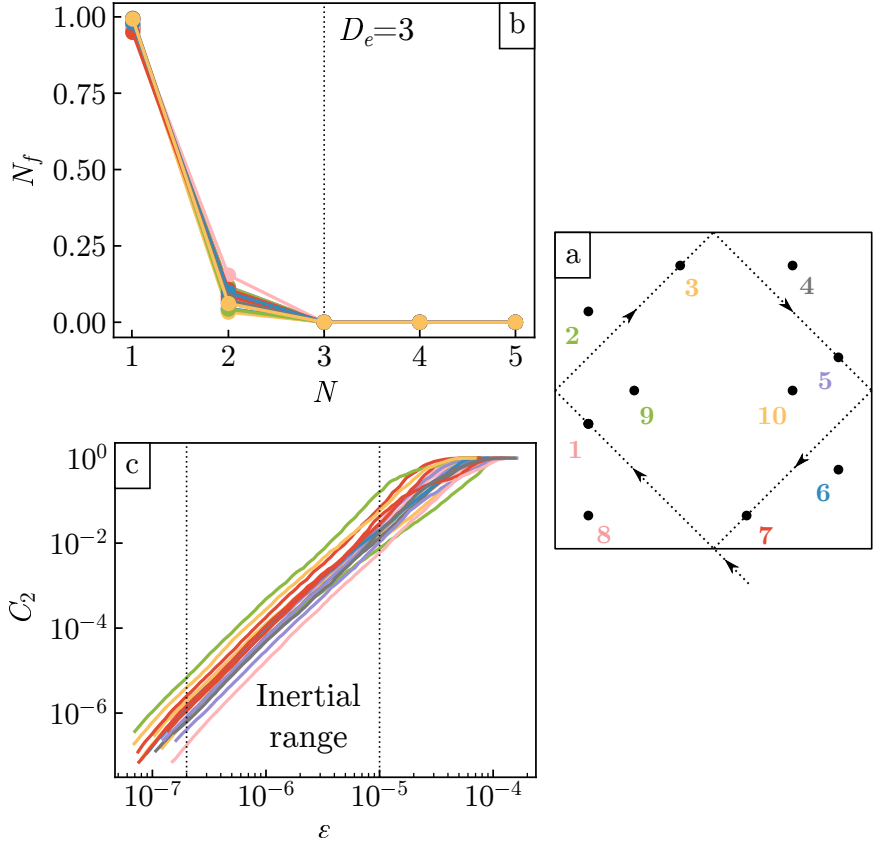}
  \caption{
    (Color online)
    False neighbours number $N_f$ and correlation sums $C_2$ for $A=2.3$ from velocity time-series taken at different positions in the cavity.
    (a) Illustration of the positions investigated in the cavity mid-height plane.
    The dashed line is the axis of the acoustic beam.
    (b) The fall to zero of the number of false neighbours indicates a global embedding dimension of $D_e=3$.
    (c) The power law exponent of $C_2$ in the inertial range gives the correlation dimension $D_2$.
    The spatial homogeneity of $D_e$ and $D_2$ in this case suggests that the whole cavity is governed by the same dynamical system.
  }
  \label{fig:NL_prop_spatial_evolution}
\end{figure}

The non-linear dynamics analysis is performed on time-series of the velocity (generally $u_t'$, as used in section \ref{sec:scen-trans-chaos}), denoted $s(t)$ from now on.
As for the power spectral densities, we checked that all quantities derived in this section are location-independent.
As an example, the correlation sum (used to compute the correlation dimension $D_2$) is plotted for $A=2.3$ and for several positions inside the cavities in figure~\ref{fig:NL_prop_spatial_evolution}.
To ensure that the signals are long enough to capture the whole dynamics of the system, the non-linear properties presented thereafter were computed for increasing signal length, until proper convergence was attained (\emph{i.e.} less than $1\%$ difference between the non-linear properties computed on a given signal and on $75\%$ of it).

\subsection{Attractor}
\label{sec:attractor-evolution}
\begin{figure}[tb]
  \centering
  \includegraphics[width=.7\columnwidth]{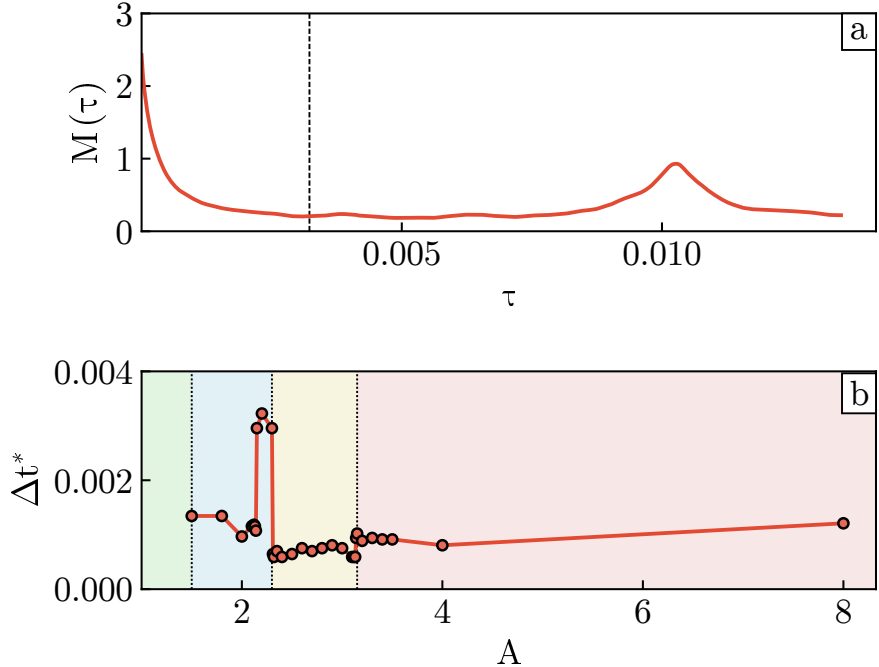}
  \caption{
    (Color online)
    (a) Example of mutual information computed from $u_t'$ for $A=2.2$.
    The first minimum is highlighted by the vertical dashed line and gives an optimal time scale $\Delta t^* \approx 0.003$.
    (b) Evolution of the optimal time scale $\Delta t^*$ with the acoustic forcing $A$.
    From left to right, the vertical dotted lines delimit the regions identified on figure \ref{fig:spectrum_evolution}: the steady region (green), the first sequence (blue), the second sequence (orange) and the regimes with an attractor of higher dimensions (red).
  }
  \label{fig:delta_t_22}
\end{figure}

We start by visualizing the time-series dynamics in the phase space using the method of delays.
The first step in achieving this is to find a timescale $\Delta t^*$ that is sufficiently short to capture the fastest timescale of the system and, at the same time, sufficiently long for the system to actually evolve between time steps.
As recommended by \citet{kantz_nonlinear_2004}, we shall define $\Delta t^*$ as the time interval that corresponds to the first minimum of the mutual information $M(\tau)$:
\begin{equation}
  M(\tau)= \sum_{ij} p_{ij}(\tau)\ln \frac{p_{ij}(\tau)}{p_i p_j},
\end{equation}
where $p_i$ is the probability to find the value of time-series in the $i$-th interval and $p_{ij}(\tau)$ the joint probability that an observation falls into the $i$-th interval and the observation after a time $\tau$ falls into the $j$-th interval.
\citet{abarbanel_analysis_1993} point out that this method provides better results for systems with a non-linear behaviour than the alternative definition of $\Delta t^*$ based on the first zero of the correlation function.
Figure~\ref{fig:delta_t_22}a shows a typical example of mutual information $M(\tau)$, while the variations of the optimal time scale $\Delta t^*$ with $A$ are shown in figure~\ref{fig:delta_t_22}b.
The high value of $\Delta t^*$ for $A \in [2.2, 2.3]$ is correlated to the appearance of low frequencies (see figure~\ref{fig:spectrum_evolution}).

\begin{figure*}
  \centering
  \includegraphics[width=\textwidth]{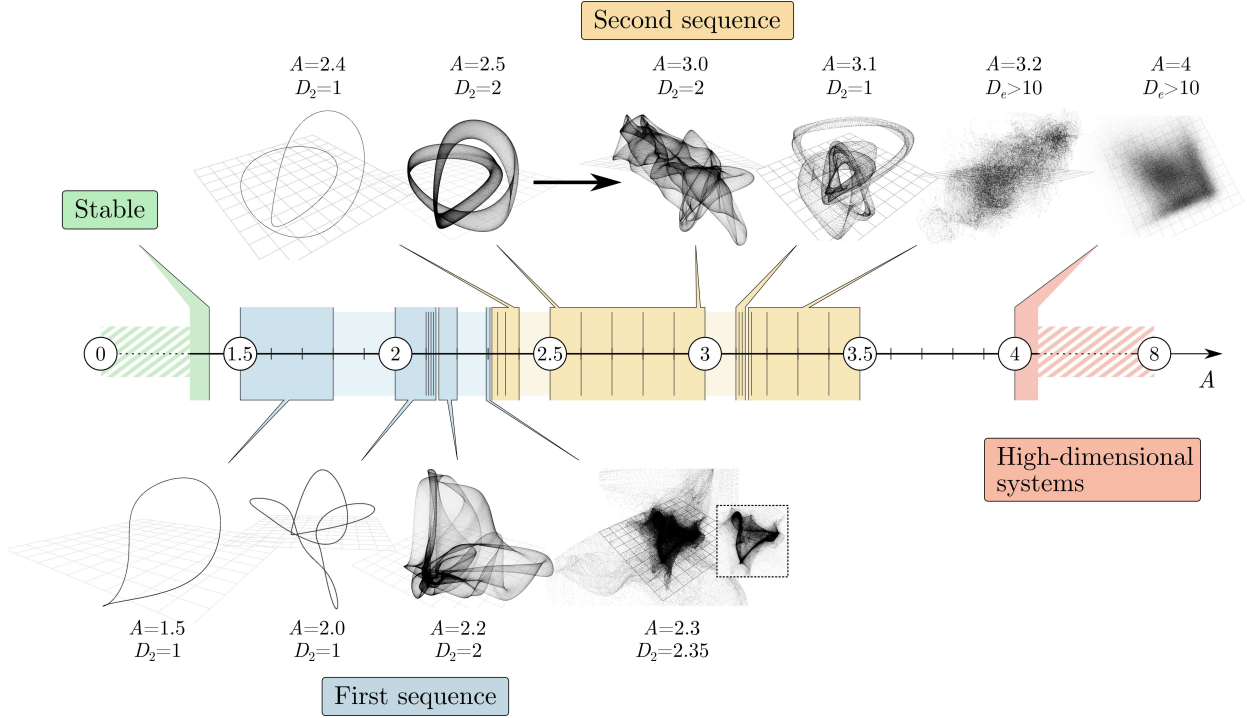}
  \caption{
    (Color online)
    Evolution of the attractors represented in three-dimensional phase spaces $\left[ s(t), s(t - \Delta t^*), s(t - 2 \Delta t^*)\right]$ for increasing acoustic forcing $A$.
    Each vertical bar represents an acoustic forcing for which non-linear properties were computed from time-series of the velocity.
    The first sequence (bottom attractors) and second sequence (top attractors) leading up to chaos are clearly visible.
    Gaps in this evolution represent regions where the dynamics fundamentally changes (as reflected by the dimensions of the attractors).
  }
  \label{fig:attracteur_transition}
\end{figure*}

Knowing $\Delta t^*$, it is then possible to build a representation of the attractor in the three-dimensional phase-space (chosen for obvious practical reasons, without prejudging of the actual dimension of the attractor), by defining the delay vector:
\begin{equation}
  \label{eq:8}
  \vec{x}(t) = \left[s(t), s(t - \Delta t^*), s(t - 2\Delta t^*)\right].
\end{equation}
Figure~\ref{fig:attracteur_transition} shows the evolution of the attractor as $A$ varies.
These representations of the attractors give a qualitative view of how the dynamics of the system evolves (as an example, a movie showing orbits spanning the attractor for $A=2.2$ is provided in the Supplemental Material in \href{movie1.mp4}{\underline{movie1}}).
The succession of regimes observed in the physical space and on the power spectral densities in section~\ref{sec:scen-trans-chaos} can again be easily traced in this representation.
In particular, the two sequences identified in figure \ref{fig:spectrum_evolution} are clearly visible here.
Both first start with an attractor of simple shape ($A=1.5$ and $A=2.4$).
Attractors then evolve into more complex shapes that nevertheless retain a legible representation in three-dimensions.
In the last phase of both sequences, the attractors end up in a shape that tells little to the naked eye ($A=2.3$ and $A=4$).

\subsection{Embedding and correlation dimensions}
\label{sec:embedd-corr-dimens}
\begin{figure}[tb]
  \centering
  \includegraphics[width=.5\columnwidth]{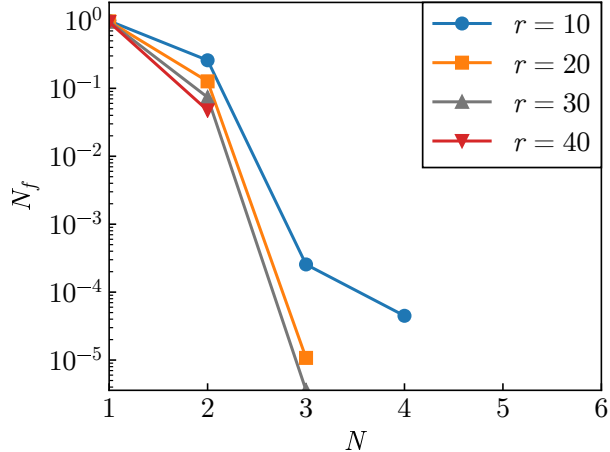}
  \caption{
    (Color online)
    Evolution of the number of false neighbors $N_f$ for increasing phase space dimension $N$ for an acoustic forcing $A=2.2$.
    Different critical divergence ratios $r$ are tested and lead to different values for the embedding dimension $D_e$.
  }
  \label{fig:false_neigh_2_2}
\end{figure}

To reconstruct the system's attractor (not necessarily lying in a three-dimensional space), we need to find its dimension, which may be or may not be an integer.
For this we must first find its \emph{embedding dimension} $D_e$, which is the dimension of the smallest linear space that contains it.
$D_e$ is found by the method of false neighbours \citep{kennel_determining_1992} which relies on the observation that if an attractor is represented in a dimension $N<D_e$, then orbits may cross each other, and points of the attractor that would be far from each other in a space of dimension $D_e$ would falsely appear as neighbours in a ``folded'' $N-$dimensional representation.
In practice, false neighbours are defined as points of the phase space that are close but for which the subsequent trajectories differ.
The embedding dimension $D_e$ is then the lowest dimension $N$ for which the number of false neighbours falls to zero.
To estimate if two trajectories differ or not, a critical divergence ratio $r$ is used.
Finding an adequate value for this ratio $r$ can be challenging, especially for chaotic attractors.
\citet{kennel_determining_1992} recommend to carry out the false neighbours analysis for a divergence ratio in the range $[10, 40]$ to ensure significant confidence in the value of $D_e$.
In this study, coherent values of $D_e$ could only be found for $r \in [20, 30]$.
To remove the uncertainty on the embedding dimension $D_e$, we performed an additional inspection of the Poincar\'e sections (see section~\ref{sec:pointc-sect-symm} where this technique is explained in more details).
A typical example of the variation of the false neighbours number $N_f$ with dimension of the embedding space $N$ is presented in figure~\ref{fig:false_neigh_2_2}.
The embedding dimension provides the size of the vector $\vec{x}_{D_e}(t)=[s(t), s(t-\Delta t^*), \ldots, s(t-(D_e-1)\Delta t^*)]$ that represents the state of the system at any given time $t$.
Once $D_e$ is known, a more precise estimate of the attractor dimension accounting for the intricacies of the system dynamics is obtained by means of the correlation dimension $D_2$ \citep{grassberger_measuring_1983}.
$D_2$ is derived from the power law exponent of the correlation sum
\begin{equation}
  C_2(D_e, \epsilon) = \frac{1}{N_{pair}} \sum_{i=D_e}^N \sum_{j=D_e}^{i-w} \Theta \left( \epsilon - ||\vec{x}_{D_e}(t_i) - \vec{x}_{D_e}(t_j)|| \right),
  \label{eq:9}
\end{equation}
where $\epsilon$ is a threshold distance, $N_{pair} = (N - D_e - w)(N - D_e - w + 1)/2$ the number of considered pairs of points, $\Theta$ is the Heaviside step function, $t_i$ is the $i^{th}$ measurement instant, and $w$ is a Theiler window to avoid considering consecutive points of a time series, that are not independent \citep{hegger_practical_1999}.
The Theiler window is determined using the first minima of the space-time separation plot, as recommended by \citet{kantz_nonlinear_2004}, and corresponds to times in the range $t=4 \times 10^{-3}$ to $1.5 \times 10^{-2}$.

\begin{figure}[tb]
  \centering
  \includegraphics[width=.8\columnwidth]{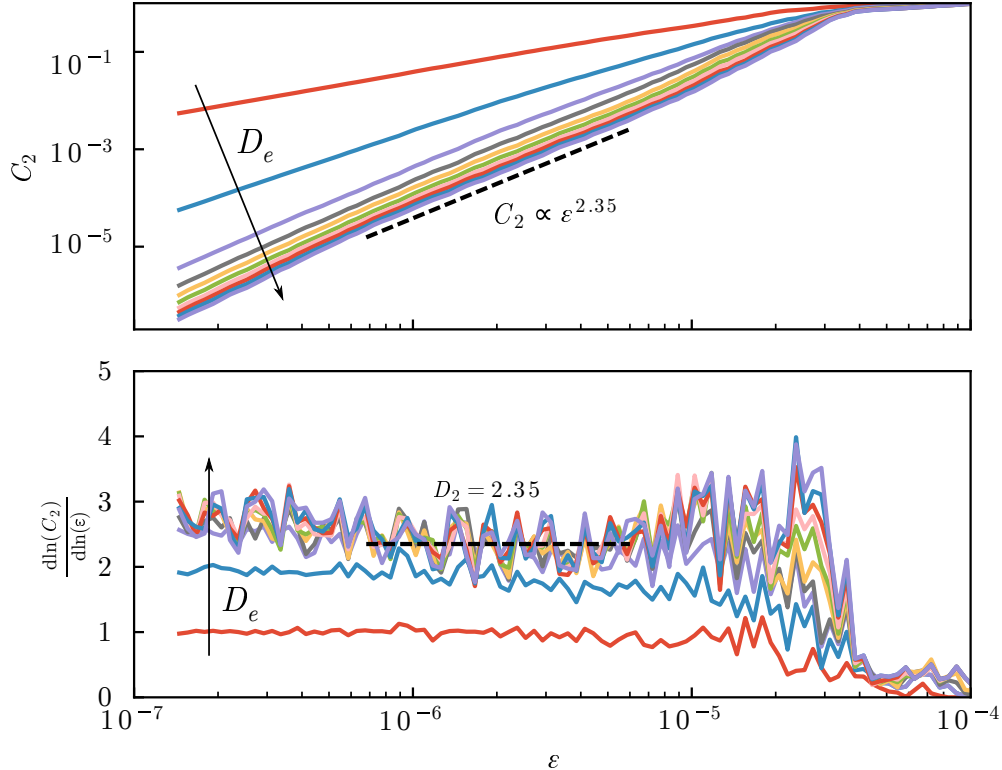}
  \caption{
    (Color online)
    Correlation sum $C_2$ computed from $u_t'$ for an acoustic forcing $A=2.3$.
    The correlation sum is computed for different embedding dimensions $D_e$ in the range from $D_e = 1$ to $10$ (represented by different colours), to ensure that it is not computed on a folded attractor.
    A fairly constant slope is observed in the scaling (central) range and indicates a fractal dimension of $D_2=2.35$.
    For small scales ($\epsilon < 7\E{-7}$), the correlation sum is known to be dominated by noise \citep{kantz_nonlinear_2004}.
    For large scales ($\epsilon > 5\E{-6}$), the self-similarity is broken by the finite extension of the attractor.
  }
  \label{fig:correlation_sum_23}
\end{figure}

\begin{figure}[tb]
  \centering
  \includegraphics[width=.8\columnwidth]{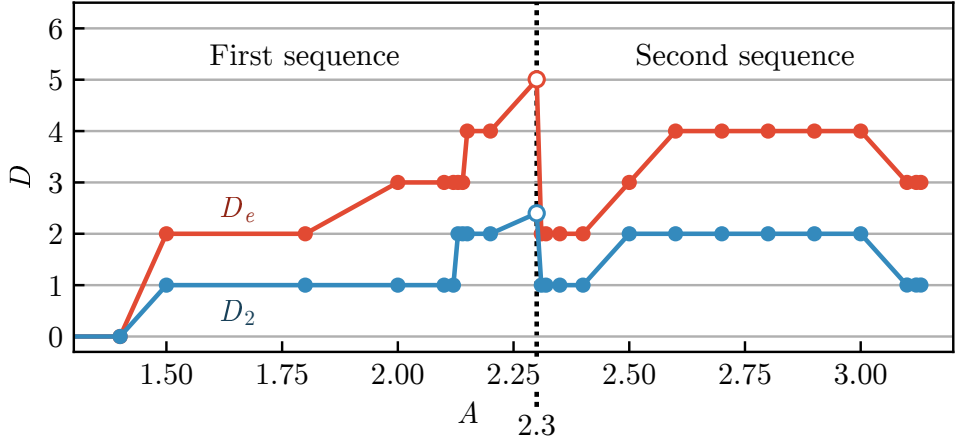}
  \caption{
    (Color online)
    Variations of the embedding and correlation dimensions ($D_e$ and $D_2$) with the acoustic forcing $A$.
    The first and second sequences are clearly visible as consecutive increases of the dimensions separated by a sudden drop.
  }
  \label{fig:nl_properties}
\end{figure}

An example of the correlation sum is plotted in figure~\ref{fig:correlation_sum_23} for $A=2.3$, and reveals a fractal correlation dimension of $D_2 = 2.35$.
The variations of both the embedding and correlation dimensions ($D_e$ and $D_2$) with $A$ are represented in figure~\ref{fig:nl_properties}.
Their variations are consistent with the evolution of the system identified in section~\ref{sec:attractor-evolution}: both dimensions increase monotonously over the interval $1.5\leq A\leq2.3$, and suddenly drop for $A=2.4$, when the dynamics brutally simplifies.
$D_e$ and $D_2$ start increasing again from $A=2.4$ to $A=3$.
The second drop in dimension between $A=3.1$ and $A=3.13$ can be explained by the frequency-locking phenomenon discussed in section~\ref{sec:char-freq}.
For $A \geq 3.15$, the dimension of the attractor increases significantly beyond the capabilities of the techniques we apply ($D_e > 10$).

The attractor for $A=2.3$ stands out as it combines a quasi-periodic behaviour and intermittent events.
This can be identified in the topology of the attractor shown in figure~\ref{fig:attracteur_transition}, where a well defined structure representing the quasi-periodic behaviour (separately presented in a dotted square) is surrounded by vast clouds of points seeded by intermittent events (See supplemental material illustrating this behaviour in \href{movie2.mp4}{\underline{movie2}}).
Unfortunately our total signal comprises only $25$ such intermittent events, and provides insufficient statistics to be able to extract the dimensions of the entire attractor.
A precise estimate would require prohibitively long numerical simulations.
Nevertheless, the attractor region corresponding to the quasi-periodic behavior is well populated and both the embedding and correlation dimensions of this region alone can be extracted by means of the methods described above.
The correlation dimension of $D_2=2.35$ is characteristic of a fractal attractor, which indicates its chaotic nature.

\subsection{Maximal Lyapunov exponent}
\label{sec:maxim-lyap-expon}
The chaotic behaviour at $A=2.3$ can be characterized further by computing the maximal Lyapunov exponent $\lambda$.
$\lambda$ is computed as the rate of spatial divergence in the phase space of two trajectories that are initially in the same neighbourhood \citep{rosenstein_practical_1993}.
We find $\lambda=7.69>0$, which, again, confirms the chaotic behaviour of the dynamical system.
The corresponding characteristic time ($T_\lambda=1/\lambda=0.13$) is of the same order of magnitude as the characteristic frequencies found in the spectral analysis in section~\ref{sec:scen-trans-chaos} ($T_a=1/f_a=0.15$).
Visual inspection of the time evolution of pairs of neighbouring points in the phase space for $A=2.3$ (See Supplemental Material in \href{movie2}{\underline{movie2}}) shows that the region of the attractor corresponding to intermittent events is a much greater source of chaos than the central part associated to a quasi-periodic behaviour.
Indeed, two initially close points remain close when in the central, dense region of the phase space, but separate very quickly when one of them undergoes an intermittent event and wanders into the outer region.

In conclusion, the non-linear analysis confirms the presence of chaos for $A=2.3$.
The fact that we were not able to compute non-linear properties for $A > 3.13$ suggests that the increase in dimensions is very abrupt at the end of the second sequence.
As a last step, we shall now analyse the particular case of $A=2.3$ in more detail.

\subsection{Poincar\'e sections and symmetry breaking}
\label{sec:pointc-sect-symm}
\begin{figure*}[tb]
  \centering
  \includegraphics[width=\textwidth]{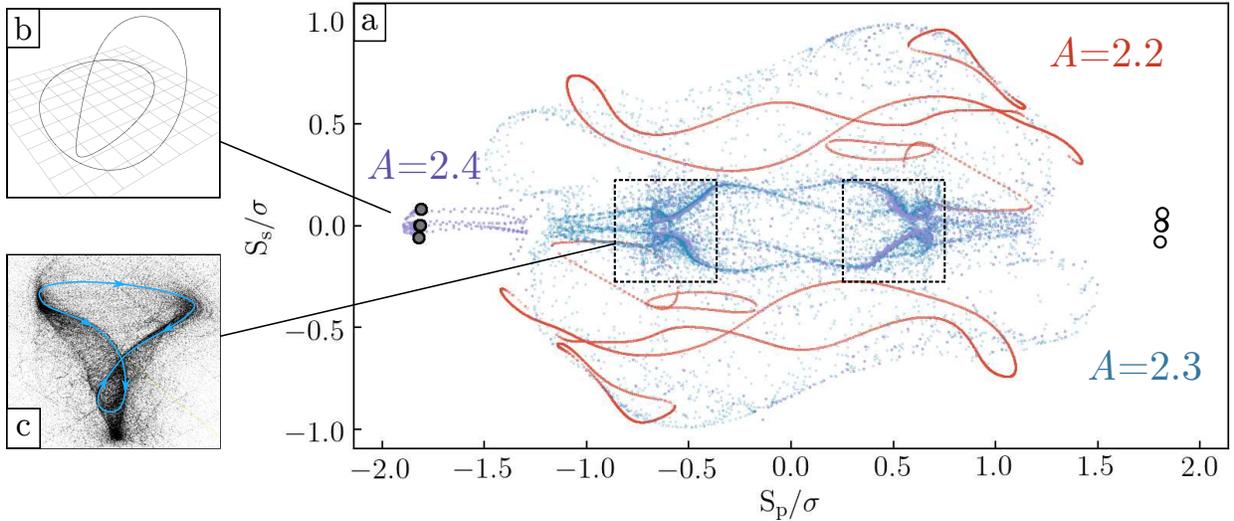}
  \caption{
    (Color online)
    (a) Comparison of the Poincar\'e section of the attractors for $A=2.2$, $A=2.3$ and $A=2.4$.
    $S_s$ and $S_p$ are the coordinates of the intersections between the Poincar\'e section and the attractors.
    The signal is normalized using $\sigma$, the standard deviation of the time-series for $A=2.2$.
    The Poincar\'e section plane is obtained using POD to ensure an optimal perpendicularity with the attractor trajectories.
    For $A=2.3$, this representation singles out the two unstable orbits (surrounded by dashed boxes), where the system remains $80\%$ of the time.
    One of these orbits is represented in (c), in the three-dimensional phase space.
    The less dense outer region for $A=2.3$ matches the region occupied by the attractor at $A=2.2$.
    For $A=2.4$, the attractor is represented with its transient part (purple points), that travels on the same Poincar\'e section as for $A=2.3$, before stabilizing (four purple circles, with two of them being juxtaposed).
    The final attractor for $A=2.4$ is represented in the three-dimensional phase space in (b).
    By symmetry, another attractor is supposed to exist for $A=2.4$, as indicated by white circles.
  }
  \label{fig:comparison_22_and_23}
\end{figure*}

The case $A=2.3$ deserves closer attention as it involves both quasi-periodic dynamics and intermittent behaviour.
To extract more information from the topology of the corresponding attractor, we shall represent its trace in Poincar\'e sections and compare it to that obtained for slightly lower and slightly higher forcing parameters ($A=2.2$ and $A=2.4$).
They are presented in figure~\ref{fig:comparison_22_and_23}.
To lay emphasis on symmetry breaking, we will use the time-series of the vertical velocity fluctuation $u_z'$ in the first jet at point $S_1$ (see figure~\ref{fig:topo_evolution}g and h), rather than the time-series of $u_t'$, used thus far.
The principal attractor directions are computed by POD (Proper Orthogonal Decomposition, see \citet{jolliffe_principal_2002} for example) to ensure that the Poincar\'e sections are plotted in planes of the phase space that are as close to perpendicular as possible to the attractor trajectories.

\begin{figure}[tb]
  \centering
  \includegraphics[width=.85\columnwidth]{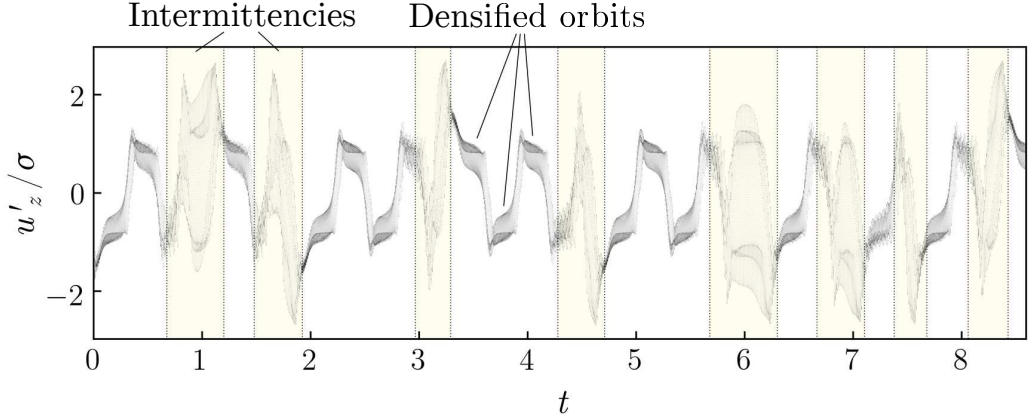}
  \caption{
    Time evolution of $u_z'$ for an acoustic forcing $A=2.3$, normalized using the standard deviation of the time-series $\sigma$.
    Vertical dotted lines delimit the regions of intermittencies.
    }
  \label{fig:23_vert_timeseries}
\end{figure}

For $A=2.2$, the Poincar\'e section is the trace of a $T^2$ torus, with the shape of a Klein bottle.
This is consistent with the embedding dimension $D_e=4$ and the correlation dimension $D_2=2$ found in the previous section.
For $A=2.3$, the Poincar\'e section exhibits two different regions that are the traces of the two regions we previously identified: (i) the outer region corresponds to intermittent events.
The topology of this region matches that of the Poincar\'e section for $A=2.2$.
(ii) Two dense regions (in dotted boxes in figure~\ref{fig:comparison_22_and_23}a) contain approximately $80\%$ of the points of the attractor.
Their shape is close to a simple closed line (figure~\ref{fig:comparison_22_and_23}c).
The system jumps periodically from one dense set of orbits to the other.
However, some of these jumps involve a long excursion into the outer parts of the attractor which corresponds to an intermittent event.
Figure~\ref{fig:23_vert_timeseries} presents the time-series of $u_z'$ for $A=2.3$, highlighting the successive switches between the densified orbits and the intermittencies.
For $A=2.4$, the system is periodic (figure~\ref{fig:comparison_22_and_23}b), which is indicated by four intersections with the Poincar\'e plane.
However, the transient part of the signal obtained for $A=2.4$ shows that the system travels for some time in the regions occupied by the attractor for $A=2.3$, before converging to a periodic attractor.

Regarding the topology of the attractors, the evolution towards a chaotic behaviour and the subsequent simplification of the dynamics follow the following scenario:
the attractor consists of a well-defined $T^2$ torus for $A=2.2$.
It loses its stability for $A=2.3$, where the dynamical system switches between three sets of unstable orbits.
The two first sets consist of dense orbits that are quasi-periodic and symmetric to each other (as shown in figures \ref{fig:comparison_22_and_23}a,c and \ref{fig:23_vert_timeseries}), while the third set is an orbit travelling through symmetric regions of the phase space that coincide with the attractor for $A=2.2$.
The system switches intermittently between the two sets of dense orbits by passing through the third one.
This behaviour is reminiscent of the Lorentz attractor, which exhibits two unstable orbits, with intermittent shifts from one to the other.
For $A=2.4$, the transient part of the signal (in purple in figure~\ref{fig:comparison_22_and_23}) shows that the system travels between the three previous sets of orbits, but finally ends up reaching a different basin of attraction, which corresponds to a state where the symmetry of the time-averaged velocity field with respect to the horizontal plane at $z=0$ is broken.

It is noteworthy that due to the up-down symmetry of the geometry and of the forcing, another periodical attractor with opposite values of $S_p/\sigma$ and $S_s/\sigma$ (represented by white points in figure~\ref{fig:comparison_22_and_23}a) is expected to exist for $A=2.4$.
Moreover, the basin of attraction leading to the attractor presented for $A=2.4$ could also be present for $A=2.3$.
In practice, however, it remains unexplored for $A=2.3$ despite the wide excursions in the phase space.
Surprisingly, the dense orbits of the configuration at $A=2.3$ share the same characteristic frequency as the attractor for $A=2.4$.
This is somewhat unexpected as both represent different areas of the phase space which co-exist (as shown by the transient part of the configuration at $A=2.4$).

To conclude this part, the chaotic behaviour of the system for $A=2.3$ is linked to the appearance of two symmetric unstable orbits.
It can be seen as an intermediate state that combines the dynamics of two non-chaotic states respectively found at slightly lower and slightly higher forcing parameters.
This is also the last calculated state (in the sense of increasing $A$) before the symmetry of the time-averaged velocity field is broken.


\subsection{High-dimensional systems}
\begin{figure}[tb]
  \centering
  \includegraphics[width=.9\textwidth]{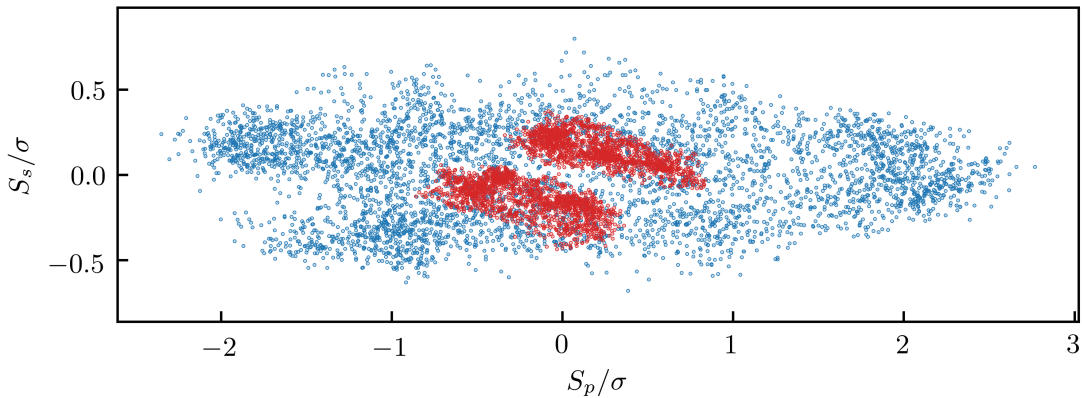}
  \caption{
    (Color online)
    Poincar\'e sections for a high-dimensional system ($A=3.2$).
    The blue section is obtained from a local quantity (transverse velocity at the jet).
    The red section is obtained from a global quantity (integral of the vertical vorticity in one quarter of the horizontal plane at mid height of the lower half of the cavity).
    For both signals, high frequencies have been filtered out at $f=1000$ using a Gaussian filter.
    A structure close to a 2D-torus is clearly visible on the red section.
  }
  \label{fig:global_quantities_3_2}
\end{figure}

The non-linear dynamics properties for acoustic forcing greater than $A=3.2$ are challenging to obtain using the method described in the previous sections because of their high dimension (see Figure~\ref{fig:attracteur_transition}).
In such a case, relying on global quantities instead of local ones, as well as filtering the high frequencies has been shown to help describe the dynamics \citep{buzug_optimal_1992, faranda_stochastic_2017}.
Having investigated several global quantities in this spirit, the integral of the vertical vorticity in quarters of horizontal planes at mid height of the upper half or lower half of the cavity came out as most useful.
Analysis based on these quantities recovered the results obtained with local times series for low-dimensional systems (up to $A=3.1$), in terms of embedding and correlation dimensions.
For high-dimensional systems, global quantities whose high frequencies have been filtered out ($f > 1000$) using a phase-preserving Gaussian filter highlighted the presence of a 2D-torus for an acoustic forcing of $A=3.2$ (Figure~\ref{fig:global_quantities_3_2}).
This indicates that the 2D-torus observed for $2.5<A<3.0$ persists at higher forcing but is concealed within the higher dimensional part of the attractor incurred by broad-band fluctuations likely associated to turbulence.
For higher acoustic forcing such as $A=4$ on the other hand, the intensity of the fluctuations becomes higher and time series analysis of the global quantities does not reveal any recognisable structure.
\section{Conclusion}
\label{sec:conclusion}
We conducted frequency and dynamical systems analyses of the acoustically driven recirculating flow obtained through successive reflections of an acoustic beam on the walls of a square cavity.
Both methods concur to show that the system can sustain states of low-dimensional chaos for specific values of the forcing parameter $A$ that measures the intensity of the acoustic forcing.
When increasing $A$, however, the systems follows a peculiar pathway to the chaotic state with a number of remarkable features:
First, the transition to chaos splits into two consecutive phases, where a succession of Hopf and period-doubling bifurcations between oscillatory states of increasing complexity leads to potentially chaotic states.
Between the end of the first phase and the beginning of the second ($2.3<A<2.4$), the dynamics of the system drastically simplifies.
This rather unexpected behaviour is consistent with former experimental observations by \citet{cambonie_flying_2017}, as are the low oscillations frequencies in the basic periodic states.
During this complex evolution, the physical states of the system depart relatively little from the basic flow topology of the steady recirculating flow, with one notable exception: in the hinge-state that immediately precedes the simplification of the dynamics, the time-averaged flow loses its up-down symmetry and fluctuations in the velocity field drop in intensity.
Second, reconstruction of the underlying attractors out of the time-series of the velocity field showed that the hinge-state was indeed chaotic (with a positive Lyapounov exponent $\lambda =7.69$ and a fractal correlation dimension of $D_2 = 2.35$).
The topology of the attractor suggests that intermittencies arise from excursions of orbits between two distinct regions of the attractor.
Each of these regions corresponds to the dynamics of the non-chaotic states respectively observed at slightly lower and slightly higher values of the forcing parameter ($A=2.2$ and $A=2.4$).
Finally, the second phase of the evolution ends up in a seemingly high-dimensional state, where the structure of the 2D torus persists, but is concealed by the appearance of high-frequency, high-dimensional fluctuations.
Along the way, a second simplification of the dynamics arises out of a frequency locking phenomenon at $A=3.1$.
The analysis of the regimes at high forcing ($A>4$) remains an open challenge.
The nature of the turbulence that may ensue may differ from that arising from the destabilisation of a classical free jet.
This could significantly impact the mixing properties of the flow.
The rich dynamics revealed in the present study raises the question of the existence of a dynamical model that would reproduce its succession of bifurcations.
Such a model would notably allow to further study the state of the system around the symmetry breaking and for high acoustic forcing.

\begin{acknowledgments}
  The authors acknowledge support from the Carnot institute Ing\'enierie@Lyon and the PHC Maghreb Partnership Program No. 36951NG.
  Support from the PMCS2I of Ecole Centrale de Lyon and the P2CHPD of University Lyon 1 for the numerical calculations is also gratefully acknowledged.
  The authors wish to thank Florence Raynal for fruitful discussions, and Amine Kass for his work as a master student.
  Alban Poth\'erat is supported by a Royal Society Wolfson Research Merit Award (Ref WM140032) and would like to express his gratitude to INSA-Lyon for the invited professor stays that made this collaboration possible.
\end{acknowledgments}

\end{document}